\documentclass[fleqn,usenatbib]{mnras}
\usepackage{anyfontsize}
\usepackage{newtxtext,newtxmath}
\usepackage{amsmath}
 
\usepackage[T1]{fontenc}
\usepackage[normalem]{ulem}

\DeclareRobustCommand{\VAN}[3]{#2}
\let\VANthebibliography\thebibliography
\def\thebibliography{\DeclareRobustCommand{\VAN}[3]{##3}\VANthebibliography}

\usepackage{graphicx}	
\usepackage{amsmath}	

\title[X-ray variability of QSOs with eROSITA]{Characterising the X-ray variability of QSOs to the highest Eddington ratios and black hole masses with eROSITA light curves}

\author[A. Georgakakis et al.]{
Antonis Georgakakis$^{1}$\thanks{e-mail: age@noa.gr},  Angel Ruiz$^{1}$, Johannes Buchner$^{2}$, Iossif Papadakis$^{3, 4}$, Maria Chira$^{1}$, 
\newauthor Kirpal Nandra$^{2}$, Shi-Jiang Chen$^{2,5}$, Maurizio Paolillo$^{6,7,8}$, 
Qingling Ni$^{2}$, Mara Salvato$^{2}$, 
\newauthor Thomas Boller$^{2}$,  Andrea Merloni$^{2}$  
\\
$^{1}$Institute for Astronomy and Astrophysics, National Observatory of Athens, V. Paulou \& I. Metaxa 11532, Greece\\
$^2$ Max Planck Institute for Extraterrestrial Physics, Giessenbachstrasse, 85741 Garching, Germany\\
$^{3}$Department of Physics and Institute of Theoretical and Computational Physics, University of Crete, 71003 Heraklion, Greece\\
$^{4}$Institute of Astrophysics, Foundation for Research and Technology, 71110 Heraklion, Greece\\
$^{5}$Department of Astronomy, University of Science and Technology of China, Hefei 230026, People's Republic of China\\
$^6$Dipartimento di Fisica "Ettore Pancini", Universit\`a Federico II, via Cinthia, 80126, Napoli, Italy\\
$^7$INAF - Osservatorio Astronomico di Capodimonte, via Moiariello 16, 80131, Napoli\\
$^8$Istituto Nazionale di Fisica Nucleare, Sezione di Napoli, I-80126 Napoli, Italy
}

\date{Accepted XXX. Received YYY; in original form ZZZ}

\pubyear{\the\year{}}

\begin{document}
\label{firstpage}
\pagerange{\pageref{firstpage}--\pageref{lastpage}}
\maketitle

\begin{abstract}
An important diagnostic of the inner structure of accretion flows onto supermassive black holes are the stochastic flux variations at X-ray wavelengths. Despite its significance, a systematic characterisation of the  statistical properties of the X-ray variability to the highest Eddington ratios and most massive black holes is still lacking. In this paper we address this issue using SRG/eROSITA 5-epoch light curves to characterise the mean X-ray variability of optically selected SDSS QSOs extending to black holes masses of $10^{10}$ solar and accretion rates close to the Eddington limit. The adopted variability statistic is the ensemble normalised excess variance, which is measured using a novel hierarchical Bayesian model ({\sc eBExVar}) tailored to the Poisson nature of the X-ray light curves. We find a clear  anti-correlation of the ensemble variability with black hole mass, extending previous results to time scales of months. This can be interpreted as evidence for an X-ray corona size and/or physical conditions that scale with black holes mass. We also find an unexpected increase of the ensemble normalised excess variance close to the Eddington limit, which is contrary to the predictions of empirical variability models. This result suggests an additional variability component for fast growing black holes that may be related to 
systematic variations of the hot corona size with Eddington ratio or shielding of the hot corona by an inner puffed-up disk and/or outflows. 
\end{abstract}




\section{Introduction}\label{sec:intro}

It is now well established that most massive spheroids in the local Universe host black holes in their nuclear regions with masses that may exceed $10^9\rm M_\odot$ \cite[e.g.][]{Kormendy_Ho2013, Mehrgan2019}. These extreme objects are believed to accumulate their mass over cosmic time primarily via accretion of material from their immediate environments \citep[e.g.][]{Soltan1982}. During this process huge amounts of energy are produced, which are observed as radiation across the electromagnetic spectrum thereby defining the astrophysical class of Active Galactic Nuclei (AGN). A fraction of the generated energy is believed to be injected into the interstellar and/or intergalactic media \citep[e.g.][]{Silk_Rees1998, Fabian1999, King2003}, thereby changing their physical conditions (e.g. temperature, pressure). This form of feedback can be sufficiently violent to impact the evolutionary path of the entire galaxy, i.e. by modulating the formation of new stars \citep[e.g.][]{Fabian2012Review, Dasyra2022}. As a result, the feeding and feedback cycle of supermassive black holes (SMBHs) has emerged as a key component in simulations and theories of the evolution of galaxies across the lifetime of the Universe \citep[e.g.][]{Somerville_Dave2015}. Despite the potential significance of the phenomenon, it remains controversial how the black hole feeding and feedback cycle operates in detail and how the released energy couples to the host galaxy medium. Addressing these questions requires understanding the formation and evolution of accretion flows as well as a quantitative description of the structure and dynamics of matter in the vicinity of the growing supermassive black holes (SMBHs). The major observational challenge for such studies is that the relevant physical processes occur on scales that are too small to be resolved spatially with either current or planned instrumentation, except from a handful of nearby systems with favourable properties \citep[e.g. the black holes of the Milky Way or the M87 galaxy,][]{EHT2022, EHT2024}.

This is where time domain astrophysics has the potential to provide a way forward. Accretion rate fluctuations \citep[e.g.][]{Lyubarskii1997, Li_Cao2008, Liu2016}, variations in the physical conditions of the accretion disk \cite[e.g.][]{Kelly2009, Dexter_Agol2011, Jiang_Blaes2020} or magnetic turbulence \citep{Sun2020} imprint temporal variations  over a wide range of time scales in the radiation produced by the feeding and feedback cycle of supermassive black holes. These can be used to set limits on the approximate size of the light emitting regions based on the causality principle or reconstruct the spatial structure of the system via reverberation mapping techniques. For example, causality indicates that the shortest variability timescales of the X-ray emission of AGN \citep[from few hundreds seconds to days,][]{Mushotzky1993, Reeves2021} correspond to an upper limit of $\la 1-30$  gravitational radii ($R_G=G\cdot M\, / c^2$) for the size of the emitting region. Moreover, measuring the time delay of variability patterns between the soft and hard X-ray emission of AGN provides further constraints on the location and size of the corona. This is because part of the radiation generated in the corona is reflected off the inner accretion disk and into the line of sight \cite[e.g.][]{Nandra_Pounds1994}. Therefore, variability in the direct coronal emission (dominant at harder X-rays) will also be mirrored in the reflected X-ray radiation at softer energies with a temporal delay that corresponds to the light travel time from the corona to the inner disk \cite[e.g.][]{Fabian2009, Uttley2014}. Such observations show that the size of this region is compact and typically less than a few gravitational radii \citep[e.g.][]{Emmanoulopoulos2011, DeMarco2013, Zoghbi2014, Kara2016, Epitropakis2016}. 

Reverberation mapping of the UV to optical continuum of AGN also provides information on the size and physical conditions of the accretion disk. Within the framework of the standard thin-disk theory \citep{Shakura_Sunyaev1973} and the assumption that the coronal flux variations are echoed by the accretion disk \citep[e.g.][]{Cackett2007}, wavelength-dependent time lags are expected in the UV/optical bands \citep[e.g.][]{Kammoun2021_model, Jha2022}. Modeling  the observed delays in the variability pattern toward longer wavelengths can therefore constrain the accretion disk size and its temperature profile \cite[e.g.][]{Edelson2015, McHardy2018, Edelson2019, Kammoun2019, Kammoun2021_obs, Kammoun2023}. If the modelling above is further extended to include predictions on the observed Power Spectral Density of the stochastic flux variations at X-ray, UV and optical wavelengths then additional constrains on the geometry and properties of the inner accretion flow are possible, e.g. the height of the corona and the Eddington ratio \citep{Panagiotou2020, Panagiotou2022}. On larger spatial scales, correlated time lags between the UV/optical and infrared continuum flux variations of AGN can constrain the properties of the dusty torus, such as its size \citep[][]{Suganuma2006, Koshida2014, Yang2020_torus, Mandal2024, Kim2024_torus} and temperature structure \citep[][]{Lyu2019_torus}.

The significance of the stochastic flux variability of AGN for understanding the inner structure of accretion flows has motivated studies to better characterize the statistical properties of these variations. A central question in that respect is the dependence of the variability on the fundamental properties of the system, such as black hole mass and Eddington ratio, to shed light on how the accretion flow changes as a function of these parameters. A statistical approach for addressing this problem that has been gaining momentum in recent years is the ensemble variability estimation. This involves the measurement of the average (ensemble) variability statistic (e.g. excess variance, \citealt{Nandra1997}; structure function, \citealt{Kozłowski2016}) using light curves or more generally repeat observations of a large population of sources selected to have similar properties (e.g. accretion luminosity, black hole mass or Eddington ratio). The underlying assumption is that the light curves of individual objects in the sample represent realisations of the same underlying variability process that one wishes to constrain. A feature of the method is that it can potentially yield interesting signal for the population even if the observations of individual objects are noisy. 

The ensemble variability estimation approach is becoming increasingly popular because of the availability of large area, nearly all sky, time domain photometric surveys particularly in the optical, e.g. the Sloan Digital Sky Survey Stripe 82 \citep{York2000}, the  Panoramic Survey Telescope and Rapid Response System \citep[Pan-STARRS,][]{Kaiser2010} or the Zwicky Transient Facility \citep[ZTF,][]{Masci2019}. These datasets are combined with large AGN samples \citep[e.g.][]{Paris2018, Lyke2020} to study how the amplitude of the flux variations at UV/optical wavelengths depends on the physical parameters of the accreting system \citep[e.g.][]{MacLeod2010, Zu2013, Simm2016, Yu2022, DeCicco2022, Arevalo2023, Petrecca2024}. The general picture emerging is that of an anti-correlation of the optical variability amplitude with black hole mass, although in detail the significance of the trend depends on the timescale of the observed variations. The variability amplitude is also shown to inversely scale with Eddington ratio and wavelength. These findings can be understood in the context of the scaling of the accretion disk size with black hole mass. This introduces a characteristic (black hole mass dependent) timescale in the system, below which the amplitude of the observed flux variations strongly depends on the timescale that the variability is measured \citep[e.g.][]{Arevalo2023}. Moreover, the disk area that contributes to the observed emission is suggested to increase toward higher Eddington ratios and longer wavelengths, thereby reducing the amplitude of the observed flux variations across a wide range of timescales \citep[][]{Panagiotou2022, Petrecca2024}.

The progress in characterising the optical variability of AGN is not mirrored at shorter X-ray wavelengths that probe the corona and the innermost regions of the accretion flow. Although it is well established that the X-ray variability amplitude at fixed time scale anti-correlates with black hole mass \citep[e.g.][]{Ponti2012, Akylas2022} the relation to Eddington ratio remains less well constrained.
 For example, measurements of the Power Spectral Density (PSD\footnote{The PSD describes the distribution of the variance of a light curve in Fourier frequencies and is often  approximated by a broken power law functional form.}) of few nearby Seyferts reveal a characteristic frequency that scales inversely with black hole mass and positively with Eddington ratio \citep[][]{McHardy2006}. Variability studies of larger low redshift AGN samples on timescales of several hours \citep[e.g.][]{Ponti2012} or several months  \citep[e.g.][]{Papadakis_Binas2024} suggest an overall decrease of the PSD amplitude with increasing $\lambda_{Edd}$. Such a dependence is also supported by temporal analysis of the X-ray photometry of large quasar populations \citep{Georgakakis2024, Prokhorenko2024} as well as modeling of the ensemble variability of X-ray selected AGN at moderate and high redshift on timescales of several days to years \citep{Paolillo2017, Georgakakis2021}.

Previous studies on the characterisation of the AGN X-ray variability are hampered by typically small sample sizes and limited dynamic range in e.g. Eddington ratio or black hole mass. This is because until recently there were no time-domain X-ray observations over large areas of the sky, similar to the numerous UV/optical surveys dedicated to transient science. This has changed however, with the start of operations of the eROSITA X-ray telescope \cite[extended ROentgen Survey with an Imaging Telescope Array,][]{Predehl2021} on board the SRG \citep[Spectrum-Roentgen-Gamma][]{Sunyaev2021} spacecraft. Since its launch the eROSITA has completed nearly 5 surveys of the X-ray sky,  thereby providing a unique data set for temporal investigations. 

In this paper we extract new eROSITA light curves of a large sample of QSOs selected in the Sloan surveys \citep{Lyke2020} to estimate their ensemble normalised excess variance \citep{Nandra1997, Paolillo2017} on rest-frame time scales of several months and explore the dependence on black hole mass and Eddington ratio.  The sample size translates to a large dynamic range for the above physical quantities and also allows exploration of new regions of the parameter space that are not accessible to previous investigations. The temporal analysis uses a new Bayesian algorithm ({\sc eBExVar}, Section \ref{sec:method}) for the estimation of the ensemble normalised excess variance that robustly accounts for upper limits in the X-ray light curves by modeling the Poisson nature of the single epoch X-ray observations. Throughout the paper we adopt a cosmology
with $H_0 = 70 \rm \,km \,s^{-1}$, $\Omega_\Lambda = 0.7$ and $\Omega_\Lambda = 0.3$.

\section{Observations}\label{sec:observations}

The AGN sample used in this work is culled from the Sloan Digital Sky Survey (SDSS) quasar catalogue data release 16 \citep[DR16Q][]{Lyke2020}, which includes 750,414 spectroscopically confirmed QSOs targeted by the Sloan telescope. Black hole masses, bolometric luminosities and Eddington ratios for the DRQ16 QSOs are from \cite{Wu_Shen2022}. We further subselect a total of 204,964 QSOs that overlap with the part of the eROSITA All Sky Survey (eRASS) whose proprietary rights lie with the German eROSITA consortium (eROSITA-DE). 

In this work we use the repeat scans of the sky by the eROSITA telescopes to construct the X-ray light curves of individual QSOs and study their ensemble variability. The eROSITA has surveyed the full sky 4 times (eRASS1 to 4) with a 6 months cadence, with an additional 5th scan (eRASS5) that is about 40\% complete. The statistical methodology adopted in this work (see Section \ref{sec:method}) to infer the ensemble variability properties of QSOs uses aperture X-ray photometry. The \texttt{apetool} task of the eROSITA Science Analysis Software System \citep[eSASS,][]{Brunner2022} is used to extract X-ray photon counts, estimate the background level and corresponding mean exposure time in the 0.2-2.3\,keV energy band at the positions of DR16Q QSOs. The adopted aperture size corresponds to the 75\% EEF (Encircled Energy Fraction) of the eROSITA PSF at the position of interest. X-ray photometry is performed separately in each of the eROSITA All Sky Surveys (eRASS1 to 5; c030 pipeline version data products) resulting in light curves with a cadence of 6\,months and duration of up to 24\,months (5 epochs).

Contamination of the aperture photometry products by photons associated with the  wings of the eROSITA PSF of nearby X-ray sources is an issue that can potentially bias the analysis. We address this by first finding the subset of DRQ16 QSOs that have eRASS1 X-ray counterparts. We use the cross-matching between the eRASS1 and the Legacy Survey \citep{Dey2019} Data Release 10 catalogues \citep{Salvato2025arXiv, Zenteno2025} to select a total of  17,765 unique DRQ16 QSOs that are securely associated with eRASS1 X-ray sources. We then define exclusion regions of 45\,arcsec around all X-ray sources with detection likelihood  {\sc det\_ml}$>7$ \citep[][]{Merloni2024}. We reject any DRQ16 QSO that overlaps with the exclusion regions and is not associated with the corresponding X-ray sources based on the \cite{Salvato2025arXiv} identification catalogue (i.e. is not one of the 17,765 DRQ16 QSOs that are securely associated with eRASS1 sources). The selections above results to 204,714 unique SDSS QSOs with eRASS1 to 5 light curves.

X-ray diffuse emission associated with the hot gas of clusters of galaxies may contaminate the aperture photometry at the positions of SDSS QSOs. We therefore remove from the sample those that lie in the close vicinity of known X-ray selected clusters (see Section 5.2 of \citealt{Merloni2024}) as explained in \cite{Georgakakis2024}. There are a total of 4061 unique SDSS QSOs that fulfill the above criterion. We further remove a total of 240 SDSS QSOs that lie within 6\,arcsec off blazars taken from the 5th edition of the Roma-BZCAT catalogue \citep{Massaro2015} and 529 QSOs that lie within 6\,arcsec off sources in the Combined Radio All-sky Targeted Eight GHz Survey \citep{Healey2007}. This is because the X-ray flux of these sources is dominated by  synchrotron radiation from relativistic electrons accelerated in radio jets directed within a small angle to the observer. The X-ray  variability of these sources is therefore  unrelated to the X-ray emitting corona. Blazars are a sub-set of the class of radio loud QSOs, the X-ray flux of which may also be dominated by emission from radio jets. We use the FIRST \citep[Faint Images of the Radio Sky at Twenty centimeters][]{Becker1995} survey and the 2nd data release of LoTSS \citep[LOFAR Two-metre Sky Survey,][]{Shimwell2017, Shimwell2022} to remove a total of 7514 radio loud QSOs following the methodology described in \cite{Georgakakis2024}. Finally we remove Broad Absorption Line QSO canidates based on the BALnicity index of the C\,IV\,$\lambda1549$ region listed in the DRQ16 catalogue,  {\sc bl\_civ>0}. This is because the X-ray flux variability of these sources may be dominated by absorption associated with winds and outflows launched by the accretion process.  We caution that there are overlaps among the different classes of sources discussed above.

\begin{figure}
	\includegraphics[width=\columnwidth]{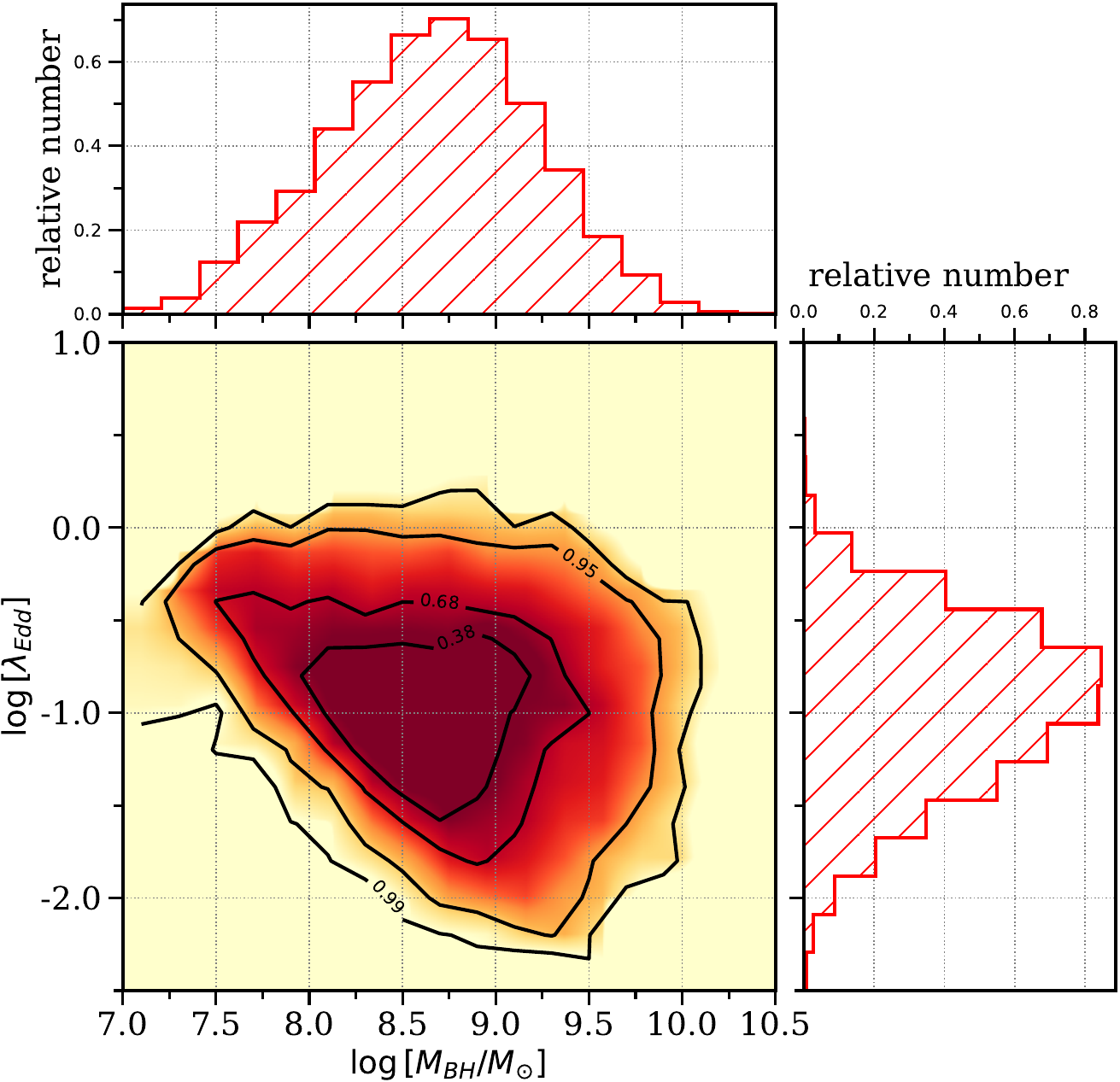}
    \caption{Distribution of the baseline sample of SDSS DRQ16 QSOs (see Section \ref{sec:observations}) on the Eddington ratio vs black hole mass plane. The contour levels are chosen to enclose 34, 68, 95 and 99 per cent of the QSO population. The 1-dimensional projections of this distribution along the black hole mass and Eddington ratio axes are also shown by the histograms on the top and to the right of the main panel respectively.}
    \label{fig:MBHLEDD-dist}
\end{figure}

\begin{figure}
	\includegraphics[width=\columnwidth]{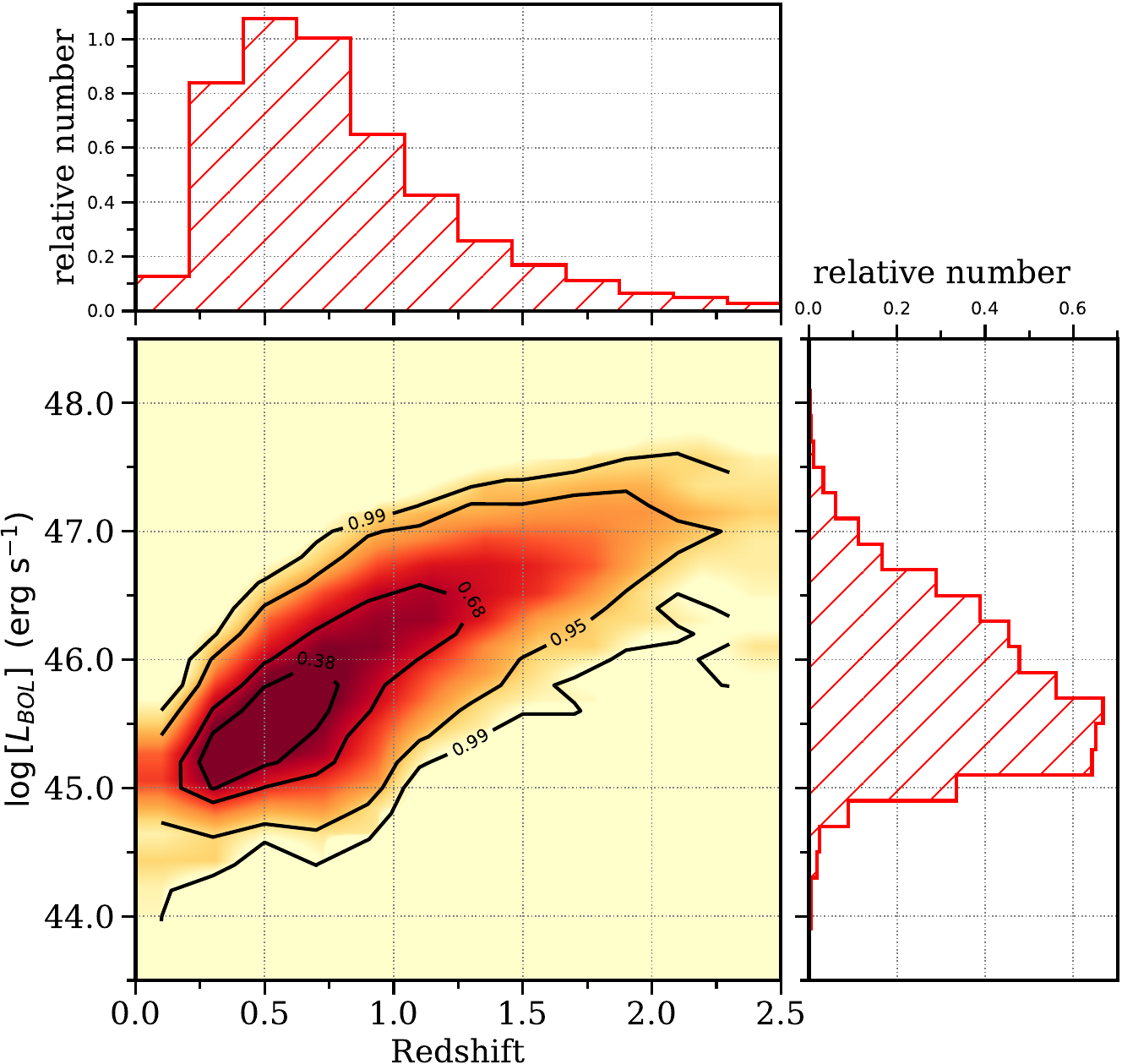}
    \caption{Distribution of the baseline sample of SDSS DRQ16 QSOs (see Section \ref{sec:observations}) on the bolometric luminosity vs redshift plane. The contour levels are chosen to enclose 34, 68, 95 and 99 per cent of the QSO population.
    The 1-dimensional projections of this distribution along the bolometric luminosity and redshift axes are also shown by the histograms on the right and at the top of the main panel respectively.}
    \label{fig:LBOLZ-dist}
\end{figure}

\begin{figure}
	\includegraphics[width=\columnwidth]{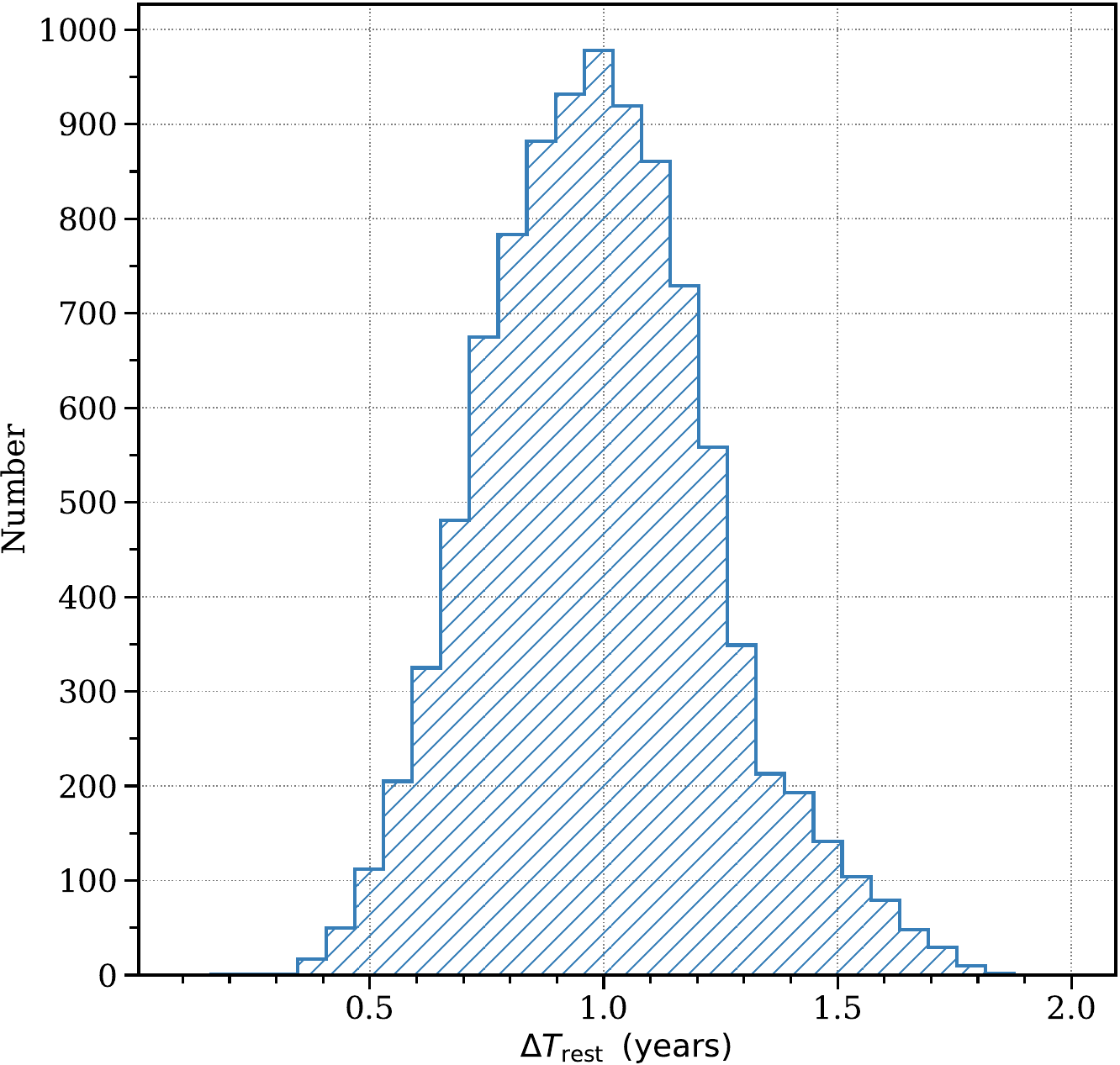}
    \caption{Histogram of the rest-frame light-curve duration probed by the baseline sample of SDSS DRQ16 QSOs (see Section \protect\ref{sec:observations}).}
    \label{fig:DT}
\end{figure}

Following the selection criteria above, the sample of SDSS DRQ16 QSOs consists of a total of 186,241 unique sources. A large fraction of the light curves have very low number of counts per epoch that provide limited information on their variability. We therefore choose to analyse only those light curves that have a minimum of 10 photon counts in at least a single eROSITA epoch. The sensitivity of the results and conclusions on this threshold is discussed in Section \ref{sec:results}. The final (baseline) sample used in this paper to determine the ensemble variability of SDSS DRQ16 QSOs consists of a total of 9678 sources, of which 6030, 
3648 have 4-epoch (eRASS1 to eRASS4) and 5-epoch  (eRASS1 to eRASS5) light curves respectively, with a cadence of 6\,months. Figure \ref{fig:MBHLEDD-dist} shows the distribution of the baseline subsample  on the Eddington ratio vs black hole mass plane. In Figure \ref{fig:LBOLZ-dist} we present the distribution of the sample on the bolometric luminosity vs redshift plane. Figure \ref{fig:DT} plots the distribution of rest-frame light curve durations of the  baseline subsample. It shows that our analysis probes timescales of about 1\,yr at the rest-frame of the source. 

\section{Ensemble Normalised Excess Variance estimation}
\label{sec:method}

The normalised (intrinsic) variance (NIV or $\sigma^2_{NIV}$) of a light curve with vanishing measurement errors estimates the intrinsic flux variability of a source and is defined as 

\begin{equation}\label{eq:niv}
    \sigma^2_{NIV} = \frac{1}{(N-1) \cdot \overline{f_X}^2}\,\sum_{1}^{N} \left( f_{X,i} - \overline{f_X} \right)^2,
\end{equation}

\noindent where the summation is over the number of photometric data points, $N$, of the light curve, $f_{X,i}$ is the X-ray flux at the epoch $i$ and $\overline{f_X}$ is the mean X-ray flux of the source. Photometric observations also include noise and therefore the contribution of random flux uncertainties to the estimated variance must be accounted for. This is done by modifying Equation \ref{eq:niv} to include the shot-noise variance of each of the photometric observations of the light curve, $\sigma^2_i$. This leads to the definition of the normalised excess variance (NEV or   $\sigma^2_{NEV}$) 

\begin{equation}\label{eq:nev}
    \sigma^2_{NEV} = \frac{1}{(N-1)\cdot \overline{f_X}^2}\,\sum_{1}^{N} \left[ \left( f_{X,i} - \overline{f_X} \right)^2 -\sigma^2_{i} \right].
\end{equation}

\noindent \cite{Almaini2000} argue that the relation above is a maximum likelihood estimator of the intrinsic light-curve variance only in the case of normally distributed flux errors. \cite{Allevato2013}  further demonstrate that, as long as biases due to the red-noise nature of the variability process are taken into account, Equation \ref{eq:nev} provides an unbiased estimate of the excess variance irrespective of the sampling pattern and the signal-to-noise of the light curves. However, for extreme sparse sampling and/or signal-to-noise ratio values less than 3, Equation \ref{eq:nev} is not useful, because the inferred NEV distribution becomes very broad and highly asymmetric (although the mean is still consistent with the intrinsic excess variance). Instead they proposed the use of ensemble estimates, based on the use of light curves of many points, which have better statistical properties.

In this section we explore alternative approaches for measuring the NEV that correctly account for the Poisson nature of X-ray light curves. Indeed, typical X-ray sources in any given observation have limited photon counts that can be described as a Poisson process. In this case the $\sigma^2_{i}$ term of Equation \ref{eq:nev}  depends on the instantaneous flux, $f_{X,i}$, of the source. A measurement of $f_{X,i}$ from the photometric observations themselves already includes the random uncertainty that ones wishes to account for. Bayesian hierarchical models are proposed to account for this limitation \citep[e.g.][]{Buchner2022, Bogensberger2024}. Suppose, for example, a source with a light curve consisting of $N$ observations at epochs $t_i$, with the index $i$ between 1 and $N$. Let us further assume that at the epoch $t_i$ the source has $N_{i}$ photon counts within an aperture with Encircled Energy Fraction $EEF_{i}$, the background level is $B_{i}$ and the exposure time at the source position is $t_{i}$. The probability that the instantaneous source flux at $t_i$ is $f_{i}$ can then be expressed by the Poisson probability

\begin{equation}\label{eq:prob-epoch-i}
    P(f_{i} \,|\, O_{i}) = \frac{e^{-\lambda_{i}}\cdot\,\lambda_{i}^{N_{i}}}{N_{i}!},
\end{equation}

\noindent  where $O_i$ signifies the observations at hand for the time epoch $i$ of the light curve (i.e. $N_{i}$, $B_{i}$ etc). The Poisson expectation value $\lambda_{i}$ is

\begin{equation}\label{eq:expectation-i}
    \lambda_{i}=f_{i}\cdot ECF_{i}\cdot t_{i}\cdot EEF_{i}\,+B_{i}.
\end{equation}

\noindent $ECF_{i}$ is the energy to flux conversion factor and depends on the source's spectral model and the characteristics of the X-ray telescope/detector that observed it. The next step is to link the single epoch fluxes $f_{i}$ by assuming that they are drawn from a log-normal distribution with mean $\mu$ and scatter parameter $\sigma$. 

Under the above conditions the likelihood of the light curve for the source under consideration is the product of the $N$ Poisson probabilities given by Equation \ref{eq:prob-epoch-i} and the probability that the source has a given mean flux and intrinsic variance

\begin{equation}\label{eq:like-single}
\mathcal{L} = \prod_{i=1}^{N} P(f_{i} \,|\, O_{i} ) \cdot \mathcal{Y} \left( f_{i} \,| \,\mu,\; \sigma \right), 
\end{equation}

\noindent where $\mathcal{Y}$ signifies the log-normal probability of observing the instantaneous flux $f_i$ given the parameters $\mu$, $\sigma$. The first is related to the median source flux and the second is an estimator of the intrinsic variance of the noisy light curve, i.e. can be used to determine the NEV. Quantitatively, the median flux of the source is given by ${\rm e}^{\mu}$ and the normalised variance of the light curve is related to the $\sigma$ parameter of the log-normal distribution as

\begin{equation}\label{eq:log-norm-nxsv}
     \sigma^2_{NEV} = {\rm e}^{\sigma^2} - 1.
\end{equation}

\noindent Bayesian inference using the likelihood of Equation \ref{eq:like-single} provides an estimate of $\sigma$ from Poisson data and hence, measures the normalised excess variance via Equation \ref{eq:log-norm-nxsv}. The likelihood above has already been proposed by \cite{Buchner2022} for finding intrinsically variable sources and by \cite{Bogensberger2024} and \cite{Boller2025} for inferring the normalised excess variance of X-ray light curves. 

Next we expand the equations above to the ensemble normalised variance of a population. This is the case of the simultaneous modeling of the light curves of many sources that are assumed to have the same (or similar) underlying intrinsic variance. This approach enables inference on the variability properties of a sample of sources, which individually may have noisy light curves. Although the information content of each of them may be limited, when considered together they may provide interesting constraints. For a sample with size $N_{\rm src}$ the likelihood of all the light curves is the product of the likelihoods of individual sources (e.g. Equation \ref{eq:like-single}) with additional terms that describe the probability distribution function of the intrinsic variance at the population level. We assume that the excess variance of individual sources is drawn from a log-normal distribution with mean and scatter parameters $\varSigma$ and $B$ respectively. Under these conditions Equation \ref{eq:like-single} can be rewritten as

\begin{equation}\label{eq:like-ensemble}
\begin{split}
\mathcal{L} = \prod_{k=1}^{N_{src}} 
\prod_{i=1}^{N} & P(f_{i, k} \,|\, O_{i, k} ) 
\cdot \\
& \mathcal{Y} \left( f_{i, k} \,| \,\mu_k,\; \sigma_k \right)
\cdot \\
& \mathcal{Y} \left(\sigma^2_{NEV, k} \,| \, \varSigma,\; B \right). 
\end{split}
\end{equation}

\noindent The index $k$ runs over the number of sources in the sample, while the index $i$ loops through the photometric data points of the light curves of individual sources. $P(f_{i, k} \,|\, O_{i, k})$ is the Poisson probability that the flux of the source $k$ with photometric observations $O_{i, k}$ (e.g. counts, exposure time, background level) at epoch $i$ is $f_{i, k}$. The flux of the source $k$ is assumed to follow a log-normal distribution, $\mathcal{Y} ( f_{i, k} \,| \,\mu_k,\; \sigma_k )$, with parameters $\mu_k$, $\sigma_k$. The first one is related to the mean/median flux of the source. The  $\sigma_k$ parameter is an estimator of the intrinsic variance of the noisy light curve of the source $k$ and hence, can be used to approximate the NEV via Equation  \ref{eq:log-norm-nxsv}. The second log-normal in Equation \ref{eq:like-ensemble} with parameters $\varSigma$, $B$ describes the probability distribution function from which the normalised variances of the population are drawn. For $N_{src}$ light curve realisations of a red-noise process the corresponding variances follow a distribution that depends on the shape of the underlying power spectrum and resembles a $\chi^2$ with low effective degrees of freedom \citep{Vaughan2003, Allevato2013}. Such a distribution has a tail that extends to high variances. For our application we approximate this distribution with a log-normal $\mathcal{Y} \left(\varSigma,\; B \right)$ that links the variances of individual sources to the same underlying distribution. This assumption is confirmed by the validation simulations presented in Appendix \ref{ap:validation}. Applying Equation \ref{eq:like-ensemble} to light curves with Poisson noise provides an estimate of the parameters $\varSigma$, $B$. The ensemble normalised excess variance of the population is then estimated as  $\sigma^2_{NEV} = {\rm e}^{\varSigma}$, i.e. the median of the log-normal distribution $\mathcal{Y} \left(\varSigma,\; B \right)$. The new  hierarchical Bayesian methodology for estimating the ensemble NEV of populations is referred to as {\sc eBExVar} ({\sc e}nsemble Bayesian E{\sc x}cess V{\sc ar}iance). In Appendix \ref{ap:validation} we describe simulations for testing {\sc eBExVar}.
 
\section{Results}\label{sec:results}

The observed light curves of individual sources consist of a list of X-ray photometric observations (aperture counts, expected background level, mean exposure time) at the different eROSITA epochs. For each light curve we determine the time interval of each epoch relative to the first observation in the series. These time intervals are then scaled to rest frame by dividing with the factor (1+$z$), where $z$ is the redshift of the source under consideration. These shifted light curves are grouped into bins of black hole mass and Eddington ratio and then {\sc eBExVar} is applied to each sub-sample to determine the corresponding ensemble normalised excess variance. We choose to use a bin size of 0.5\,dex for each of the two physical parameters, which is comparable to the systematic uncertainty affecting the determination of black hole masses from single epoch optical spectra. The Hamiltonian Markov Chain Monte Carlo code Stan\footnote{https://mc-stan.org} is used to sample the likelihood of Equation \ref{eq:like-ensemble} and produce parameter posterior distributions. The free parameters of the model include (i) the mean ($\varSigma$) and scatter ($B$) of the log-normal distribution describing the normalised excess variance of the sample, (ii) the $\sigma^2_{NEV, k}$ of individual light curves (see Eq. \ref{eq:log-norm-nxsv}) that are drawn from the above distribution, (iii) the single epoch count-rate (or flux) of a source ($f_{i,k}$ in Equation \ref{eq:like-ensemble}) and (iv)  the mean count-rate (or flux) of a light curve ($\mu_k$ parameter in Equation \ref{eq:like-ensemble}). For the sampling of the likelihood of Equation \ref{eq:like-ensemble} we adopt uniform priors in the intervals [$-4,0.0$] for $\varSigma$ and [$0,1$] for $B$. In the case of $\log\mu_k$ we adopt a normal prior with scatter 0.2 and mean estimated by summing up the X-ray photons counts of the light curve of the $k$ DR16Q QSO as explained in Appendix \ref{ap:meanflux}. The choice of the latter prior is motivated by the simulations discussed in Appendix \ref{ap:validation}.

\begin{figure}
	\includegraphics[width=\columnwidth]{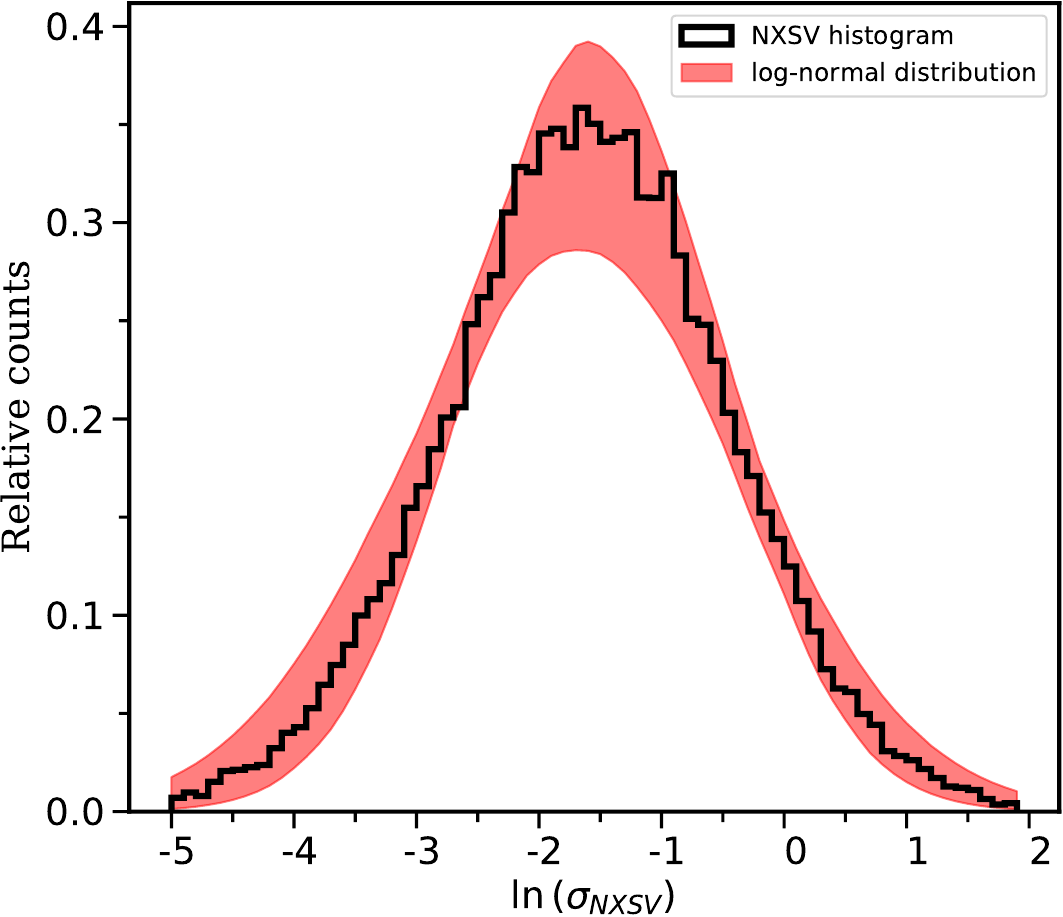}
    \caption{Inferred log-normal distribution of the normalised excess variance for the sample of SDSS DRQ16 QSOs with black hole masses in the range $\log\,M_{BH} = \rm [8.0,8.5]$ and Eddington ratios in the interval $\log\,\lambda_{Edd} = \rm [-0.5,0.0]$. The shaded region shows the 68th confidence interval of the reconstructed log-normal distribution of the population using the posteriors of the model parameters $\varSigma$, $B$ of Equation \ref{eq:like-ensemble}. The histogram is constructed from the inferred $\sigma_{NEV, k}$ (see Equation \ref{eq:like-ensemble}) of individual light curves.}
    \label{fig:hist_NXSV}
\end{figure}

Figure \ref{fig:example_LC} shows an example of a light curve that is used as input to the ensemble normalised excess variance inference code. The same figure also shows the inferred light-curve parameters produced by {\sc eBExVar} for this particular source, i.e. the single epoch and mean count-rates. Figure \ref{fig:hist_NXSV} demonstrates the ensemble variability properties produced by {\sc eBExVar}. It shows the inferred log-normal distribution describing the normalised excess variance of a QSO sample (i.e. free parameters $\varSigma$, $B$) in comparison with the inferred NEV of individual sources in the same sample.

The inferred NEV of SDSS DRQ16 QSOs in different Eddington ratio and black hole mass bins is shown in Figure \ref{fig:NXSV}. In Appendix \ref{ap:polynomial} we fit a 2nd order polynomial to the 2-dimensional space of Figure \ref{fig:NXSV} to analytically express the NEV as a function of black hole mass and Eddington ratio and facilitate the comparison of our results with future studies. For fixed Eddington ratio there is an overall trend in Figure \ref{fig:NXSV} of decreasing $\sigma_{\rm NEV}$ with increasing black hole mass, i.e. the ensemble excess variance  drops along columns from bottom to top of Figure \ref{fig:NXSV}. This trend is stronger for the two lowest Eddington ratio bins but is still evident for the subsamples with $\log \lambda_{Edd}>-1.0$. This finding extends previous results on the black-hole mass dependence of the variability amplitude of AGN from timescales of few hours \citep{Ponti2012, Akylas2022, Tortosa2023} to several months and years \citep[see also][]{Lanzuisi2014}. 

An interesting trend emerges in Figure \ref{fig:NXSV} when considering the dependence of the $\sigma^2_{\rm NEV}$ on Eddington ratio for fixed black-hole mass interval. Previous studies suggest that the amplitude of the flux variations of AGN drop with increasing Eddington ratio \cite[e.g.][]{Ponti2012, Georgakakis2024}. Instead, in Figure \ref{fig:NXSV} there a is "U"-shape dependence of the $\sigma^2_{\rm NEV}$ on $\lambda_{Edd}$. The initial decrease of the inferred excess variance from the $\log\lambda_{Edd}$ interval [-2.0, -1.5] to the one with [-1.0, -1.5] is followed by a leveling off and an increase for the highest Eddington ratio bin, [-0.5, 0.0]. This trend is stronger for the black hole mass bins with $\log M_{BH}/M_\odot<9.0$. For the most massive black holes in the sample, $\log M_{BH}/M_\odot$=[9.5, 10.0], the NEV decreases at the highest Eddington ratio bin. We caution however, that this is the sample with the lowest number of QSOs,  a total of 46.

\begin{figure*}
	\includegraphics[width=2\columnwidth]{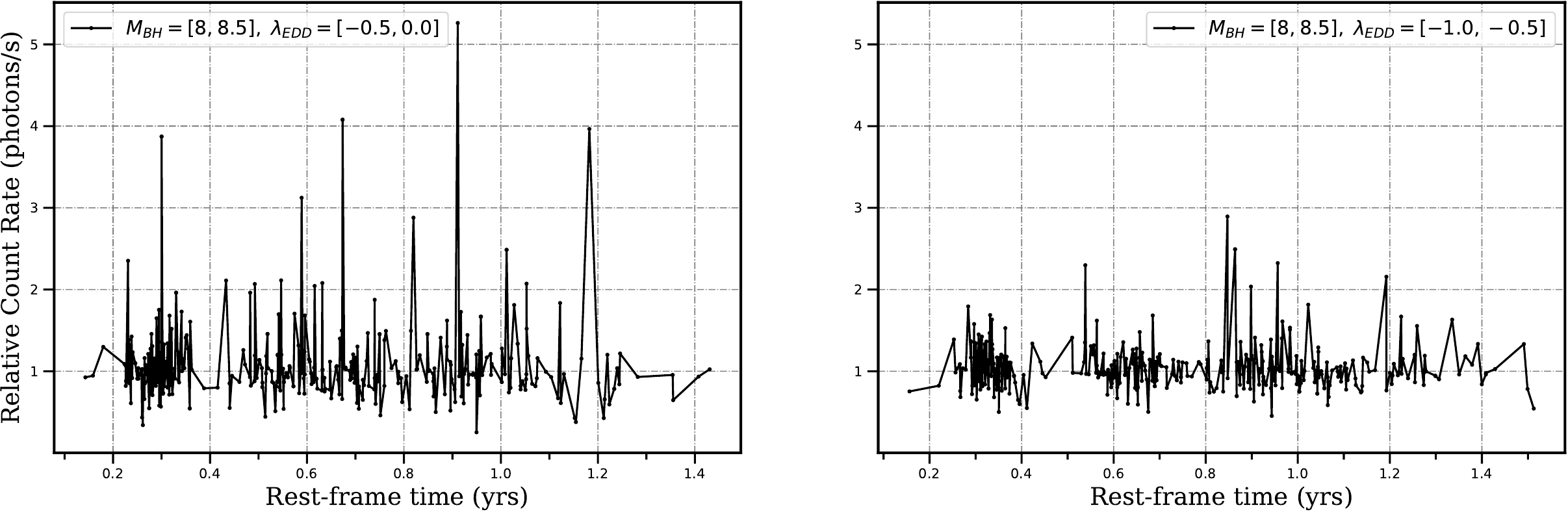}
    \caption{Visualisation of the higher level of variability of QSOs close to the Eddington limit. We compare the synthesized X-ray light curves of DR16Q QSOs selected to have black holes masses in the interval $\log\,M_{BH} = \rm [8.0,8.5]$ and Eddington ratios in the range $\log\,\lambda_{Edd} = \rm [-1.0,-0.5]$ (right hand panel) and $\log\,\lambda_{Edd} = \rm [-0.5,0.0]$ (left-hand panel). The synthesized light curves are constructed by randomly sampling data points (total of 400 in this realisation) from the light curves of individual sources shifted to rest-frame and re-scaled to a mean count rate of unity (see text for details). The high Eddington ratio QSOs on the left show higher level of variability compared to those on the right.}
    \label{fig:LCs}
\end{figure*}

It is interesting to visualise the higher variability amplitude in the case of high Eddington sources by constructing combined light curves from the individual ones drawn from the sample. For each QSO the Bayesian inference provides an estimate of the instantaneous count rate at a given epoch and the mean of the light curve. We use these values to normalise all light curves to a mean count rate of unity. We then shift each light curve by setting its first epoch to zero years and scale the subsequent time intervals to rest-frame by dividing with the factor (1+$z$). This approach enables the superposition of the individual light curves of the sample under the assumption that they have similar statistical properties. Figure  \ref{fig:LCs} shows the results of this approach for two sub-samples with black holes masses $\log\,M_{BH} = \rm [8.0,8.5]$ and Eddington ratios in the range $\log\,\lambda_{Edd} = \rm [-1.0,-0.5]$ and $\log\,\lambda_{Edd} = \rm [-0.5,0.0]$. The light curves shown in this Figure are constructed by randomly drawing a total of 400 normalised/scaled count rates (see above) from the light curves of the QSOs in each of the two sub-samples. It can be seen that the higher Eddington sub-sample ($\log\,\lambda_{Edd} = \rm [-0.5,0.0]$) demonstrates more erratic behaviour.

We also explore how empirical models for the PSD of AGN compare against the measured excess variances plotted in Figure \ref{fig:NXSV}. In this exercise  we use the fact that the NEV is an estimator of the band-limited power \cite[$NIV_{\infty}$,][]{Bogensberger2024}, i.e. the  integral of the PSD between the frequencies of interest. We adopt a PSD analytic model that follows a bending power-law  form based on the variability properties of local Seyferts \citep{McHardy2004, Gonzalez-Martin_Vaughan2012} but also motivated by NEV studies at higher redshift \citep[][]{Paolillo2023}

\begin{equation}\label{eq:psd}
PSD(\nu) = A\,\nu^{-1} \, \left(1+    \frac{\nu}{\nu_b}\right) ^{-1},
\end{equation}

\noindent where $\nu$ is the frequency. The amplitude, $A$, and break frequency, $\nu_b$, depend on the physical parameters of the accreting system, e.g. the mass of the black hole and its Eddington ratio. For the break frequency in particular we adopt the relation

\begin{equation}\label{eq:model1-nub}
\nu_b = \frac{580}{M_{\rm BH}/M_\odot} \; (\rm s^{-1}).
\end{equation}

\noindent  This parametrisation is motivated by the observations of \cite{Papadakis2004} and  \cite{Gonzalez-Martin_Vaughan2012}.  Additionally, following observational results on local Seyferts  \citep[e.g.][]{Ponti2012}, we assume that the amplitude of the PSD scales with Eddington ratio as

\begin{equation}\label{eq:model3-amp}
A = 2\cdot\nu_b \cdot PSD(\nu_b) = 3\times10^{-2}\cdot\lambda_{Edd}^{-0.8}.
\end{equation}

\begin{figure}
	\includegraphics[width=\columnwidth]{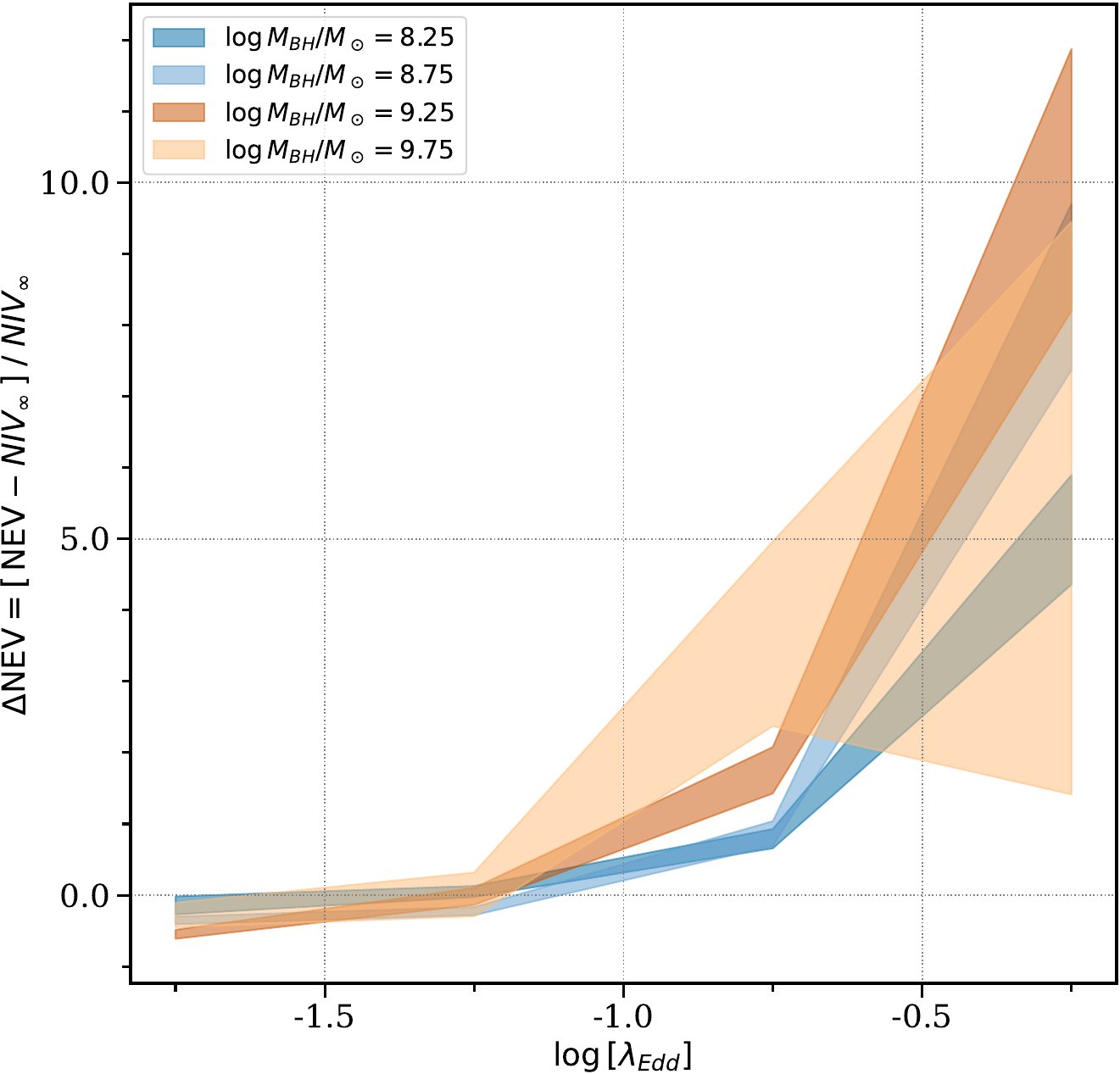}
    \caption{ Plotted as a function of Eddington ratio is the fractional difference between the observational derived NEV of DRQ16 QSOs and the prediction ($NIV_{\infty}$) of the analytic PSD model parametrised in Equations \ref{eq:psd}-\ref{eq:model3-amp}. The shaded regions correspond to the black hole mass intervals [8.0, 8.5] (dark blue), [8.5, 9.0] (cyan), [9.0,9.5] (dark brown) and [9.5, 10] (light brown). The width of the shaded regions correspond to the $1\sigma$ uncertainties (68th confidence interval around the median of the NEV posterior distribution).}
    \label{fig:DNEV}
\end{figure}

\noindent The PSD parametrisation above is used to predict the $NIV_{\infty}$ within the black hole mass and Eddington ratio bins of Figure \ref{fig:NXSV}. For a given QSO with $M_{BH}$ and $\lambda_{Edd}$ we estimate the break frequency and amplitude of its assumed PSD via Equations \ref{eq:model1-nub} and \ref{eq:model3-amp} respectively. We then integrate Equation \ref{eq:psd} between the minimum and maximum rest-frame frequencies for the QSO under consideration. The former corresponds to the rest-frame duration of the light curve, while for the latter we scale by $1/(1+z)$ the time interval of 24\,hours. This is how long it typically takes SRG/eROSITA to accumulate the total exposure time per sky position for a  single eRASS. The justification for this latter timescale is presented in Appendix \ref{ap:niv}. For each  black hole mass and Eddington ratio bin of Figure \ref{fig:NXSV} we estimate the mean $NIV_{\infty}$ of the population. This is then plotted as a function black hole mass and Eddington ratio in the top and right panels of Figure \ref{fig:NXSV}. The model predicts a decreasing  $NIV_{\infty}$ with either increasing black hole mass or increasing Eddington ratio. The latter prediction is at odds with the observations for $\lambda_{Edd}\ga-0.5$. This discrepancy is further demonstrated in Figure \ref{fig:DNEV} that plots as a function of Eddington ratio the fractional difference between the observational measured NEV and the model predicted $NIV_{\infty}$ (i.e. $\Delta {\rm NEV} = \frac{{\rm NEV} - NIV_{\infty}}{NIV_{\infty}}$) for the DRQ16 subsamples of fixed black hole mass. The quantity $\Delta {\rm NEV}$ deviates to positive values for $\lambda_{Edd}\ga-1$, i.e. the observed NEV is larger than the $NIV_{\infty}$ predicted by the empirical PSD model of Equations \ref{eq:psd}-\ref{eq:model3-amp}.
 
In Appendix \ref{ap:validation2} we explore the sensitivity of the ensemble NEV results of Figure \ref{fig:NXSV} on the definition of the DRQ16 sample. It is demonstrated that the trends with Eddington ratio and black hole mass shown in Figure \ref{fig:NXSV} are robust and do not depend on how the QSO population is selected or the adopted X-ray band. We have also repeated the analysis by relaxing the adopted threshold for the minimum number of photons in at least one light curve epoch for a DRQ16 QSO to be included in the variability sample. Using a minimum of 5 instead of 10 photons (see Section \ref{sec:observations}) results in a sample of 36,729 unique light curves, of which 11,218 with 5-epoch photometry. The estimated ensemble NEV as a function of black hole mass and Eddington ratio for this new sample is consistent within the errors with the values shown in Figure \ref{fig:NXSV}.

\begin{figure}
	\includegraphics[width=\columnwidth]{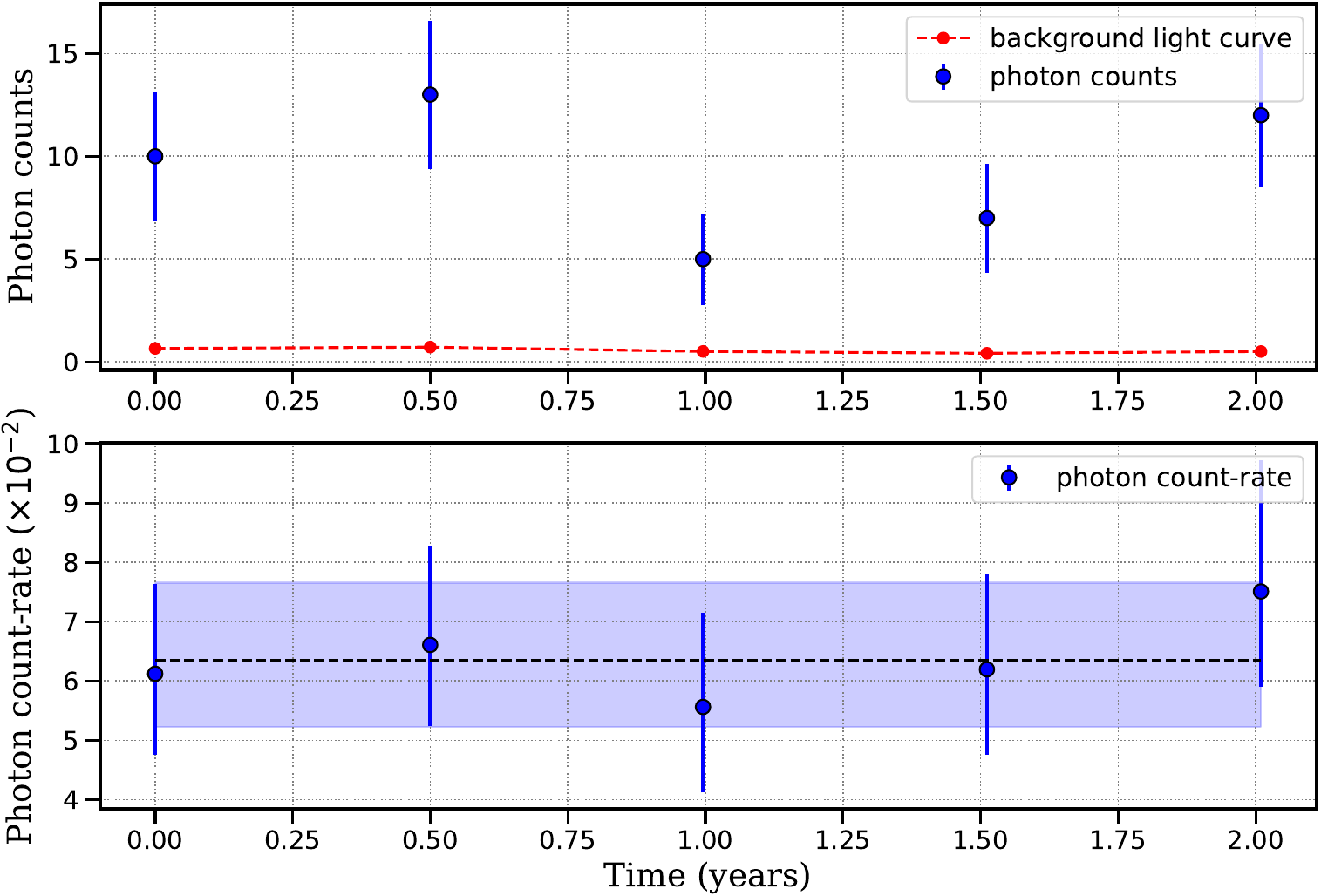}
    \caption{Example light curve of the SDSS DRQ16 QSO 022040.01-084545.3. The top panel plots the observed eROSITA photon counts (blue filled circles) as a function of the observation epoch in years. The errorbars associated with each point are Poisson uncertainties. The bottom panel plots the estimated photon count rate of the same source as function of time in years. The single epoch rates are inferred from  {\sc eBExVar} (see Section \ref{sec:method}, parameter $f_{i}$ in Equation \ref{eq:like-single} or $f_{i,k}$ in Equation \ref{eq:like-ensemble}). The uncertainties correspond to the 68\% confidence interval around the median of the posterior distributions of the single epoch photon count rates. The dashed black line shows the mean count rate of the light curve also inferred from the Bayesian approach of Section \ref{sec:method} (i.e. parameter $\mu$ in Equation \ref{eq:like-single} or $\mu_k$ in Equation \ref{eq:like-ensemble}). The shaded region is the $1\,\sigma$ uncertainty of the mean count rate.}
    \label{fig:example_LC}
\end{figure}

\begin{figure}
	\includegraphics[width=\columnwidth]{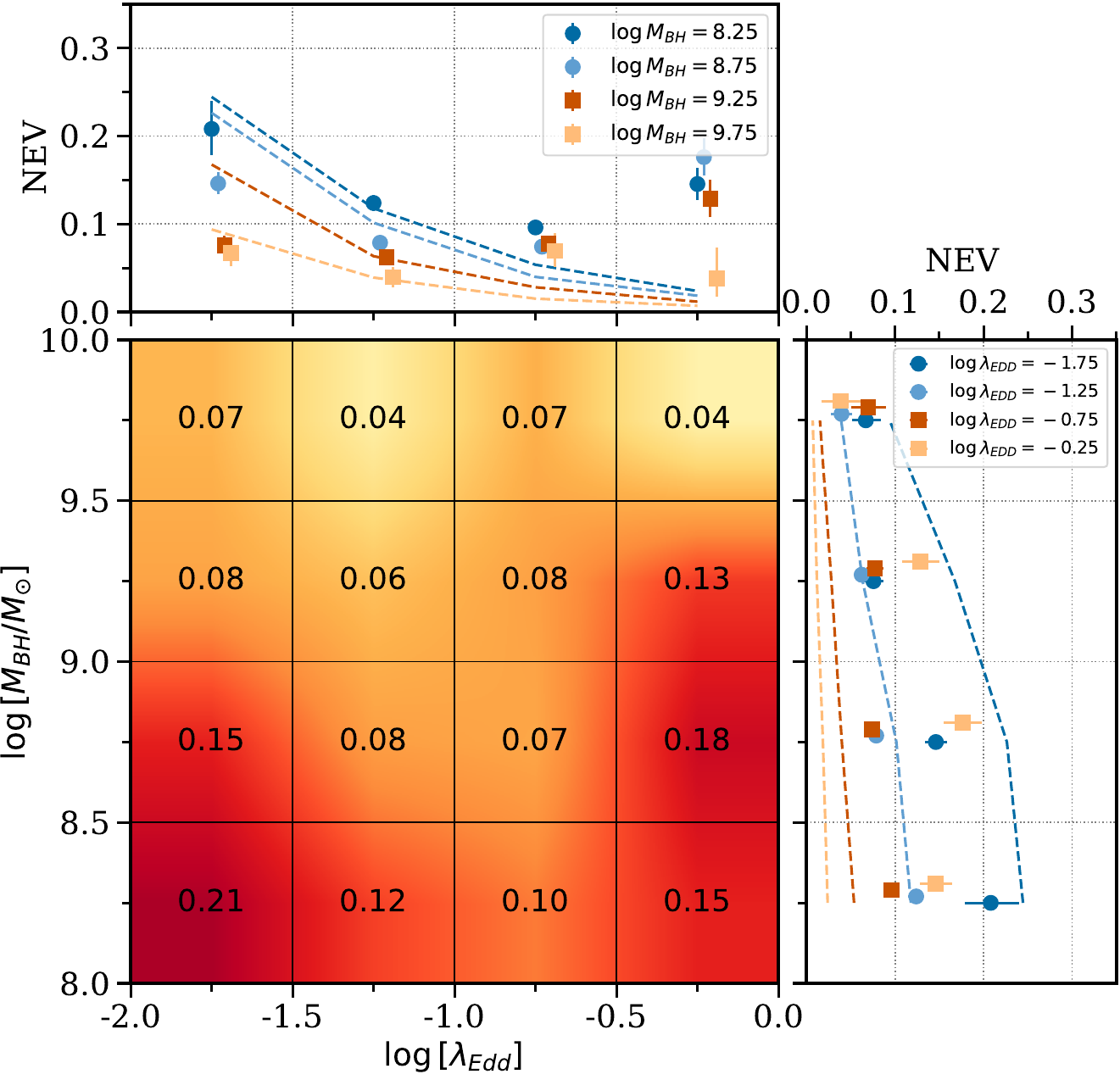}
    \caption{Ensemble normalised excess variance of DRQ16 QSOs as a function of black hole mass and Eddington ratio. The vertical and horizontal solid lines in the main panel show how the parameter space is divided in black hole mass and Eddington ratio bins with logarithmic widths of 0.5\,dex. The value at the middle of each box is the estimated ensemble normalised excess variance for DRQ16 QSOs within the corresponding black hole mass and Eddington ratio limits.  The shading is a smoothed representation of the  $\sigma_{\rm NEV}$ variations on the 2-dimensional space of $M_{BH}$ and $\lambda_{Edd}$. Darker colours correspond to higher values of  $\sigma_{\rm NEV}$. The panel on the right shows how the $\sigma_{\rm NEV}$ varies with black hole mass at fixed Eddington ratio. The legend shows the color coding and symbols adopted for the  Eddington ratio intervals, i.e. [-2, -1.5] (dark blue circles), [-1.5, -1.0] (cyan circles), [-1.0,-0.5] (dark brown squares) and [-0.5, 0] (light brown squares). The top panel shows how the $\sigma_{\rm NEV}$ varies with Eddington ratio at fixed black hole mass. The legend shows the color coding and symbols adopted for the various black hole mass intervals, i.e. [8.0, 8.5] (dark blue circles), [8.5, 9.0] (cyan circles), [9.0,9.5] (dark brown squares) and [9.5, 10] (light brown squares). The curves plotted in the top and right panels are the predictions of a parametric PSD model that depends on both Eddington ratio and black hole mass (Equations \ref{eq:psd}-\ref{eq:model3-amp}, see text for details). The color coding of the curves corresponds to different Eddington ratio (right panel) or black hole mass (top panel) intervals as indicated in the legend. The errorbars in the top and right panels correspond to the $1\sigma$ uncertainties (68th confidence interval around the median of the NEV posterior distribution).}
    \label{fig:NXSV}
\end{figure}

\section{Discussion}

The X-ray variability of AGN provides important information on the inner structure of the accretion flow  onto supermassive black holes as well as the interplay between the X-ray emitting hot corona and other components of the flow such as the accretion disk or winds launched by the AGN. In this paper we characterise the dependence of the X-ray variability on the black hole mass and the Eddington ratio of the central engine using  {\sc eBExVar}, a new hierarchical Bayesian model for the determination of the ensemble normalised excess variance of QSO populations. Features of this tool include the robust statistical treatment of the Poisson nature of X-ray observations and the homogeneous accounting of both X-ray upper limits and detections. We apply the methodology to new X-ray light curves of optically selected SDSS QSOs based on multi-epoch observations from the eROSITA All Sky Surveys. The sample is split into black hole mass and Eddington ratio bins to investigate variations of the NEV with these parameters. 

\subsection{NEV dependence on black hole mass and Eddington ratio}

A first result from the analysis is the anti-correlation of the X-ray NEV with black hole mass for rest-frame time scales of several months to years. At a given Eddington ratio interval in Figure \ref{fig:NXSV} the NEV overall decreases toward higher black hole masses. This finding extends previous X-ray studies that report similar trends but on much shorter timescales that correspond to a fraction of day \citep[e.g.][]{Ponti2012, Akylas2022, Tortosa2023}. An anti-correlation with black hole mass for temporal intervals extending to months and years is also reported for the UV/optical variability of QSOs \citep[e.g.][]{Arevalo2023, Petrecca2024}. It is  generally expected that the characteristic sizes of different accretion flow components (e.g. disk, corona) scale with the mass of the central black hole \citep{Morgan2010, Reis_Miller2013, Dovciak_Done2016}. In this context any instabilities will take longer to propagate through the system in the case of more massive black holes. Indeed, the characteristic temporal scales of the accretion flow \citep[e.g. dynamical, viscous, thermal;][]{Noda_Done2018} are proportional to black hole mass or equivalently the corresponding frequencies vary as $M_{BH}^{-1}$. As a consequence, the power spectrum of the flux variations (assumed to be a power-law, e.g. Equation \ref{eq:psd}) is expected to shift to lower frequencies with increasing black hole mass. This translates to a decrease of the variability amplitude and the observed NEV at fixed timescale for more massive black holes. 

Next we turn to the Eddington ratio dependence of the NEV in Figure \ref{fig:NXSV}. Focusing first on the two lowest $\lambda_{Edd}$ bins we find evidence that the NEV inversely correlates with the Eddington ratio. This is consistent with the expectation of empirical PSD models, motivated by studies of the X-ray flux variations of local Seyferts \citep{Ponti2012}, which suggest a monotonic drop of the variability amplitude toward higher $\lambda_{Edd}$  \citep[e.g.][]{ONeill2005} as in Equation \ref{eq:model3-amp}. This parametrisation predicts $NIV_\infty$ that are in fair quantitative agreement with the observed NEV (factor of $<1.5$; see top panel of Figure \ref{fig:NXSV} and Figure \ref{fig:DNEV}). This level of consistency is worth highlighting because it involves significant extrapolation of the empirical model PSD parameter space. The median black hole mass of the nearby Seyfert sample used to derive the empirical model is few times $10^7\,M_\odot$ and hence much lower (1-3 orders of magnitude) than those of DRQ16 QSOs. The origin of the observed anticorrelation of the variability amplitude with Eddington ratio is not well understood. It is suggested for example that the X-ray corona is heated by magnetic flares which are generated as the magnetic field of the accretion disk is amplified by the differential rotation, then rise up buoyantly to the surface where the lines reconnect and release energy \cite[e.g.][]{Haardt1994, Poutanen_Fabian1999, Hopkins2024}. Assuming that $\lambda_f$ is the rate of flares on the disk, then the normalised excess variance of the corona emission is expected to scale as $\propto 1/\sqrt{\lambda_f}$. If $\lambda_f$ is proportional to the Eddington ratio, i.e. fast accretion flows are more efficient in producing magnetic flares per unit time, then one would expect the X-ray NEV to anticorrelate with this quantity \cite[see also][]{ONeill2005}.  
Such a model would also predict that the level of X-ray 
variability continues to decrease toward the highest Eddington 
ratio bins of our sample. Interestingly, such a monotonic trend is 
not supported by the observational results shown in Figures 
\ref{fig:NXSV} and \ref{fig:DNEV}. Instead we find that at fixed 
black hole mass the X-ray variability amplitude levels off and/or 
increases for $\log \lambda_{Edd}>-1.0$. This trend is more 
striking for black hole masses in the range $\log M_{BH}/
M_\odot=8-9$, where the NEV shows a "U"-shape behaviour. It 
initially decreases with increasing Eddington ratio, but then the 
trend inverses, with the NEV rising toward the highest $
\lambda_{Edd}$ interval used in our analysis, [-0.5, 0.0]. Such a 
dependence is not reported by previous studies on the X-ray NEV of 
AGN \citep[e.g.][]{Ponti2012, Paolillo2023, Papadakis_Binas2024}, 
likely because the corresponding samples do not probe the high 
Eddington ratio regime with sufficient statistics.  It is also 
informative to compare the measured variability at X-rays with 
results at different wavebands.  At UV/optical in particular, the 
NEV of QSOs is shown to anti-correlate with Eddington ratio out to 
$\lambda_{Edd}\approx1$ on rest-frame time scales from months to 
about a year \citep{Arevalo2023, Petrecca2024}, in contrast to the 
results in Figures \ref{fig:NXSV} and \ref{fig:DNEV} on similar 
timescales. The difference in AGN variability properties as a 
function of wavelength likely provides important information on 
the structure of the accretion flow on small scales close to the X-
ray emitting corona in the case of fast growing black holes. 

Overall the evidence above points to an new variability component that is emerging for QSOs approaching the Eddington limit and is not predicted by established empirical PSD parametrisations of the stochastic X-ray flux variations \citep[e.g.][]{Ponti2012}. Although the mechanism that produces the additional variance is not clear, some progress can be made by examining classes of sources that are thought to include fast accreting supermassive black holes.

\subsection{Comparison with known high Eddington ratio AGN/QSO samples}

Narrow line Seyfert 1s \citep[NLSy1;][]{Osterbrock_Pogge1985, Goodrich1989} represent AGN with moderate mass black holes ($\log M_{BH}/M_{\odot} \la  7$) that typically accrete at a relatively high rate, approaching the Eddington limit. Although their X-ray NEV properties are, on average, similar to other AGN populations, once their moderate black hole masses are accounted for \citep[e.g.][]{Ponti2012}, they do include a fraction of highly variable sources \citep[e.g.][]{Gonzalez-Martin_Vaughan2012}. \cite{Boller2021} argue for example, that the extreme X-ray flux variations of the NLSy1 1H\,0707-495 are consistent with changes in the covering fraction of absorbing clouds intercepting the line-of-sight. Moreover, the absorber is likely part of an outflow produced by the accretion process. A different picture is proposed however, to explain flaring events in the NLSy1 MRK\,335 \citep{Grupe2012, Gallo2013, Wilkins2015, Gallo2019}. It is argued that the sudden increase in the X-ray flux of this system can be explained as beamed emission from a radially expanding corona caught in the act of launching an outflow.The evidence above underlines the importance of winds for understanding the X-ray properties of at least some NLSy1. An alternative scenario for the X-ray spectral variability of NLSy1 is that of relativistic reflection \cite[e.g.][]{Ross_Fabian2007, Fabian2012}. In this picture the X-ray corona is close to the black hole and as a result the X-rays are relativistically bent away from the observer and toward the accretion disk, where they can be reflected into the line--of--sight. X-ray flux variations in this case case are associated with changes in the fraction of the reflected radiation, which in turn may be  associated with the ionisation state of the refractor \cite[][]{Miniutti2012} or the corona height and physical conditions.

The high Eddington ratio SDSS QSOs in our sample may represent the more massive and luminous counterparts of nearby NLSy1. In this picture winds or relativistic reflection may be responsible for their observed X-ray variability properties. It is also interesting that in many highly X-ray variable NLSy1s \citep[e.g.][]{Bachev2009, Gallo2011,  Miniutti2012} the UV/optical emission appears to be stable, suggesting that the process responsible for the X-ray modulations does not affect the accretion disk emission. This trend is similar to that of QSOs in our sample, the UV/optical variability of which monotonically decreases with Eddington ratio  \citep{Arevalo2023, Petrecca2024}, in contrast to their X-ray NEV shown in Figure \ref{fig:NXSV}. 

Serendipitous X-ray surveys and monitoring campaigns in the last few years have enabled the discovery of a small yet increasing number of distant QSOs that demonstrate exceptional levels of X-ray flux variability \cite[e.g][]{Liu2019, Ni2020, Liu2022, Huang2023, Yu2023}. These QSOs tend to have high Eddington ratios, while their optical/UV continua or emission-lines are typically stable and do not appear to follow the X-ray flux variations. The emerging physical picture to interpret the observed properties of these sources is that of variable X-ray obscuration as a result of winds driven by the accretion process or shielding by the geometrically thick inner accretion disk. Our analysis of the variability properties of SDSS QSOs demonstrates that the above anecdotal results on selected QSOs are part of a systematic trend, whereby the overall variability of the population increases toward higher $\lambda_{Edd}$.

\subsection{Physical models for the X-ray variability of high Eddington ratio QSOs}

Next we explore different models for the accretion flow that could reproduce an increasing QSO X-ray variability toward high Eddington ratios. 

Simulations suggest that at high accretion rates relative to the Eddington limit the inner disc becomes geometrically thick, while the strong radiation field of the system drives dense and possible clumpy outflows \citep[e.g.][]{Ohsuga2011, Takeuchi2013, Jiang2014, Jiang2019}. These properties are conducive to high levels of X-ray variability because of changing levels of the line-of-sight absorption from either the puffed-up inner disk (e.g. changes in disc thickness) and/or the clumpy wind launched in the close vicinity of the supermassive black hole \cite[e.g. variations in cloud covering fraction and/or ionization level;][]{Boller2021, Ni2018, Ni2020}. In this picture the UV/optical radiation is likely less affected since it is produced further out in the accretion disc. The X-ray variability timescales for the picture above are challenging to predict. Assuming that highly variable NLSy1s (e.g. 1H 0707-495, \citealt{Boller2021};  IRAS 13224-3809, \citealt{Chiang2014}) represent examples of AGN that follow the model above, then one expects to observe strong variability over a wide range of timescales from days to several years, i.e. similar to the rest-frame timescales of Figure \ref{fig:DT}.

Alternatively, the higher X-ray variability amplitude close to the Eddington limit may indicate a smaller perhaps less stable X-ray emitting corona for fast accreting black holes. X-ray spectral analysis of QSOs using physically motivated models of the accretion flow structure \citep{Kubota_Done2018} indeed suggest that the hot corona radius drops with increasing Eddington ratio \citep[][]{Mitchell2023, Chen2025}.  The power spectrum of the flux variations is therefore expected to shift to higher frequencies for higher values of the Eddington ratio. This translates to a increase in the  variability amplitude and the observed NEV at fixed timescale for fast accreting black holes.

Models of the inner accretion flow onto SMBHs propose the presence of a warm ($kT\rm \sim1\,keV$) and optically thick ($\tau \sim 10-20$) Comptonization region (warm corona) on top of the standard accretion disk to explain the shape of the AGN SEDs in the extreme UV and soft X-ray regime \citep[e.g.][]{Done2012, Rozanska2015, Petrucci2018, Kubota_Done2018}.  The warm corona is thought to produce the soft-excess X-ray spectral component of AGN, which dominates below about 2\,keV \citep[e.g.][]{Petrucci2018} and is shown to become increasingly stronger for higher Eddington ratio AGN \citep[][]{Chen2025}.   However, this component is shown to be less variable than the power-law continuum in local Seyferts \citep[e.g.][]{Vaughan_Fabian2004, Parker2020, Igo2020}. It is therefore unlikely to be related to the higher variability amplitude we observe in the eROSITA 0.2-2.3\,keV spectral band for QSOs close to the Eddington limit.

Correlations between X-ray spectral changes and flux variations could test the scenarios above. For example, flatter X-ray spectra during low flux states could be evidence for the obscuration or shielding picture \citep[e.g.][]{Boller2021}. Extending ensemble variability studies to X-ray energies beyond $2\,\rm keV$ \citep[e.g.][]{Papadakis_Binas2024} would also provide important information on the origin of the observed variations. For example, lower variability amplitudes at hard relative to soft X-rays are expected in the case of the obscuration scenario \citep[e.g.][]{Miniutti2012, Fabian2012}. The drop of sensitivity of eROSITA for energies higher than about 2\,keV renders variability studies at hard X-rays challenging. {\it XMM-Newton} or, in the future, the Athena X-ray observatory are more appropriate for such investigations. The archive of {\it XMM-Newton} in particular, currently contains potentially useful data on the ensemble variability of QSOs on time scales of several hours to days. Combining this information with the results of the current paper could provide constraints on the (mean) shape of the X-ray power spectrum as a function of black hole mass and Eddington ratio. Jointly modelling the X-ray and UV power spectra of QSOs \citep[e.g.][]{Panagiotou2022,Hagen2024}  could provide further information on the interplay between accretion disk and corona on different time scales.  Finally, it is also interesting to explore links between the variability of high Eddington-ratio QSOs and the peculiar population of broad-line (optical type-I) and X-ray absorbed AGN that emerge at highest accretion luminosities \citep{Merloni2014}. The discrepant optical and X-ray classifications of these systems may be because of dust-free gas clouds within the broad-line region  \citep{Merloni2014}, possibly pointing to AGN winds.


\section{Conclusions}

This paper applies a new hierarchical Bayesian model to the eROSITA X-ray (0.2-2.3\,keV) light curves of SDSS DRQ16 QSOs to measure the ensemble normalised excess variance of the population as a function of black hole mass and Eddington ratio. Compared to previous X-ray variability studies we extend the parameter space to AGN with the most massive black holes and the highest Eddington ratios. 

First we find that the X-ray ensemble nomalised excess variance on timescales of several months anti-correlates with black hole mass at fixed Eddington ratio. This is consistent with previous studies that find similar trends but on much shorter timescales of several hours. A possible interpretation could be the scaling of the corona size and/or physical conditions (e.g. temperature, optical depth) with the mass of the compact object. 

Our analysis also suggests a complex "U"-shape dependence of the X-ray ensemble NEV on Eddington ratio. At fixed black holes mass the NEV initially decreases with increasing Eddington ratio, as expected by empirical models, but then levels off and increases again for accretion rates approaching the Eddington limit. This finding points to a additional  variability component for fast accreting black holes. 

Shielding or partial covering of the X-ray emission by either a geometrically thick disk  (expected in simulations of high Eddington ratio AGN) or outflowing material launched by the accretion process (also common in fast growing black holes) are feasible options to interpret the observations. An alternative scenario is that of a smaller size hot corona toward higher accretion rates, as suggested by recent physical models of the accretion flow.

\section*{Acknowledgements}
The authors thank the anonymous referee for their useful comments and suggestions. The research leading to these results has received funding from the Hellenic Foundation for Research and Innovation (HFRI) project "4MOVE-U" grant agreement 2688, which is part of the programme "2nd Call for HFRI Research Projects to support Faculty Members and Researchers". This work is based on data from eROSITA, the soft X-ray instrument aboard SRG, a joint Russian-German science mission supported by the Russian Space Agency (Roskosmos), in the interests of the Russian Academy of Sciences represented by its Space Research Institute (IKI), and the Deutsches Zentrum f{\"u}r Luft- und Raumfahrt (DLR). The SRG spacecraft was built by Lavochkin Association (NPOL) and its subcontractors, and is operated by NPOL with support from the Max Planck Institute for Extraterrestrial Physics (MPE). The development and construction  of the eROSITA X-ray instrument was led by MPE, with contributions from the  Dr. Karl Remeis Observatory Bamberg \& ECAP (FAU Erlangen-Nuernberg), the University of Hamburg Observatory, the Leibniz Institute for Astrophysics Potsdam (AIP), and the Institute for Astronomy and Astrophysics of the University of T{\"u}bingen, with the support of DLR and the Max Planck Society. The Argelander Institute for Astronomy of the University of Bonn and the Ludwig Maximilians Universit{\"a}t Munich also participated in the science preparation for eROSITA. The eROSITA data shown here were processed using the eSASS/NRTA software system developed by the German eROSITA consortium. This research made use of  Astropy,\footnote{\href{www.astropy.org}{www.astropy.org}} a community-developed core Python package for Astronomy \citep{astropy:2013, astropy:2018}.

\section*{Data Availability}

The {\sc eBExVar} code used in this paper is available at GitHub \url{https://github.com/ageorgakakis/ebexvar}.



\bibliographystyle{mnras}
\bibliography{example} 


\appendix
\section{Testing the ensemble NEV hierachical model}\label{ap:validation}

\begin{figure*}
	\includegraphics[width=2.0\columnwidth]{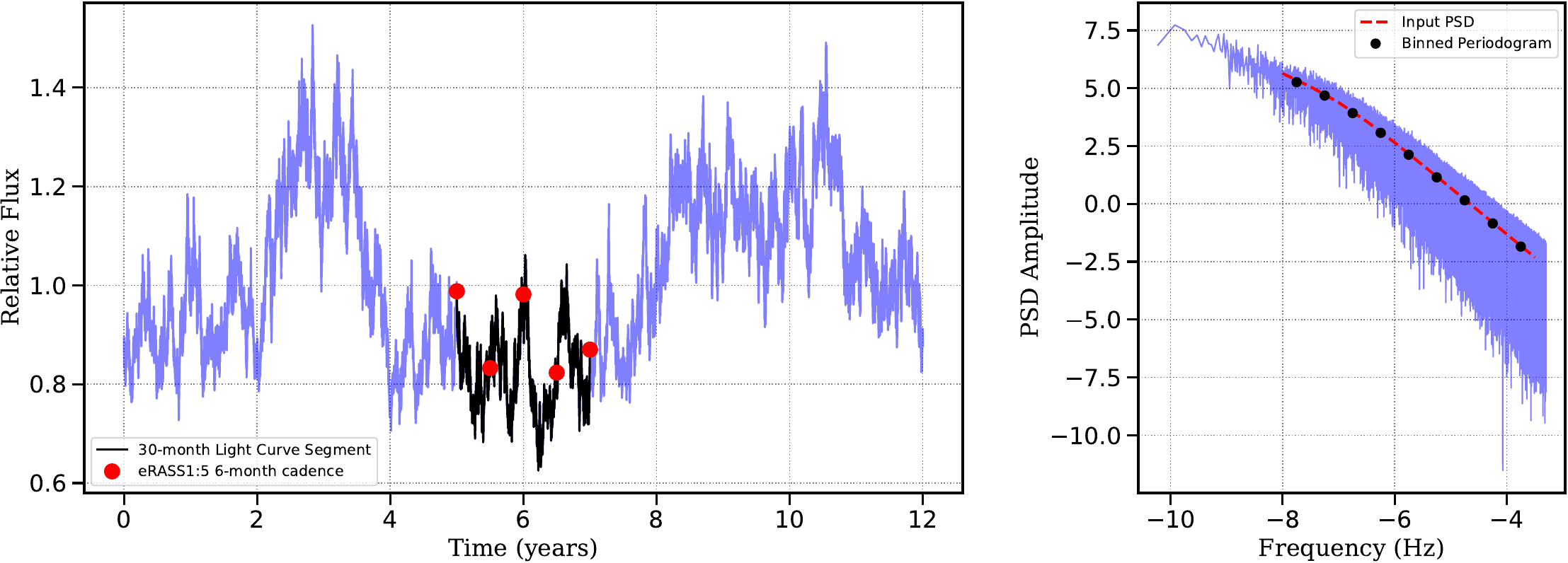}
    \caption{The left panel shows an example of a 12-year long section (blue) of a simulated light curve with a total duration of several hundred years and cadence of 1000\,s generated with the code of \protect\cite{Sartori2018, Sartori2019}. The black-shaded region of the light curve shows a segment with duration of 24 months (2\,yrs). The red circles sample the black segment with a cadence of 6 months, thereby resembling the eROSITA sky scanning frequency. The panel on the right plots the periodogram of the full simulated light curve. The black circles correspond to the mean of the periodogram in logarithmic bins of 0.5\,dex. The red line shows the input PSD that follows the broken power-law of Equation \ref{eq:niv} 
    with break frequency and amplitude given by Equations \protect\ref{eq:model1-nub}, 
    \protect\ref{eq:model3-amp} respectively. The adopted black hole mass and Eddington ratio for the PSD in these simulations are $\log M_{BH}/M_{\odot}=9.75$ and $\log\lambda_{Edd} = -0.25$.}\label{fig:lcsimulation}
\end{figure*}

In this section we develop realistic simulations of eROSITA X-ray light curves with known intrinsic variance to test the performance of  {\sc eBExVar} for estimating the ensemble normalised excess variance of AGN populations. 

\subsection{Setting up the light curve simulations}\label{ap:settingup}

The generation of light curves uses the code presented by \cite{Sartori2018, Sartori2019} based on the methods of \cite{Emmanoulopoulos2013}. We adopt a power spectral density of the flux variations that follows the bending power law of Equation \ref{eq:psd} with a break frequency that is a function of black hole mass as in Equation \ref{eq:model1-nub} and an Eddington-ratio dependent amplitude given by Equation \ref{eq:model3-amp}. The fluxes of the light curve are drawn from a log-normal distribution with zero mean and scatter of 0.5\,dex. For a given black hole mass and Eddington ratio a long light curve is produced with a total duration of few hundred years and a cadence of 1,000\,s. Figure \ref{fig:lcsimulation} presents a segment of such a light curve along with the corresponding periodogram in comparison with the input PSD model. 

Simulated light curves such as the one shown in Figure \ref{fig:lcsimulation} are used to generate observed time series of fluxes that resemble the eROSITA photon statistics and cadence. The eROSITA telescopes survey the entire sky every six months and the data used in this work includes the first 5 scans of the sky. For the simulations we therefore adopt a cadence $\rm C=6\,months$ and a total light curve duration of $\rm D=24\, months$ (2\,years). Time series with these characteristics are constructed by first randomly selecting from the full simulated light curve segments with total length of 24\,months. These are further sub-sampled with a cadence of 6 months to construct a time series of 5 epochs. Examples of a 24\,month--long segment and the corresponding 5-epoch eROSITA-like time series is over-plotted on the light curve of Figure \ref{fig:lcsimulation}. 

Next we construct eROSITA photon counts for each of the 5 epochs of the time series. The first step is to assign a mean 0.2-2.3\,keV flux $f_{X,m}$ to the $\rm D=24\, months$-long segment. As a result each of the 5-epochs, $i$, of the time series are associated with a flux $f_{X,i}$ that deviates from $f_{X,m}$ because of the intrinsic variance of the light curve. Each $f_{X,i}$ is then assigned X-ray aperture photometry values (background level, mean exposure time, encircled energy fraction of the adopted aperture) randomly drawn from the eROSITA multi-epoch photometric catalogue of SDSS DRQ16 QSOs (see Section \ref{sec:observations}). It is therefore possible to convert each flux to photon counts using the relation 

\begin{equation}
    E_{i} = f_{X, i} \cdot t_{i} \cdot ECF_{i} \cdot EEF_{i} + B_{i},
\end{equation}

\noindent where $t_{i}$, $B_{i}$, $EEF_{i}$ are the eROSITA exposure time, background value and encircled energy fraction assigned to epoch $i$. We reiterate that these values correspond to the real aperture photometry measurements of a randomly chosen SDSS DRQ16 QSO (see Section \ref{sec:observations}). $ECF_{i}$ is the energy to flux conversion factor that assumes an X-ray spectral model with $\Gamma=1.9$ and Galactic   absorption appropriate for the sky position of the SDSS DRQ16 QSO from which the background value and exposure time are drawn. Finally $E_i$ is the expectation value of the observed eROSITA photon counts. It is used to draw a Poisson deviate, $C_i$, that represents the observed integer photon counts within the aperture. The end product of this process are 5-epoch aperture photometry measurements ($C_i$, $t_i$, $B_i$, $EEF_{i}, ECF_{i}$, $i=1-5$), which can be used to determine the intrinsic variance of the segment from which they are drawn.  

Our analysis focuses on the ensemble normalised excess variance of populations rather than individual sources. The methodology described above is therefore repeated to generate $N_S$ groups of 5-epoch eROSITA-like aperture photometry measurements, where $N_S$ represents the total number of sources in the sample. Each of the $N_S$ 5-epoch time-series has a cadence of 6 months and is drawn from randomly selected segments of the same long light curve. It is this simulated dataset that is used as input to  {\sc eBExVar} (see Section \ref{sec:method}) to determine the ensemble NEV of the sample. 

For testing and validation purposes we wish to cover a broad range of intrinsic flux variances. This is achieved by generating light curves for PSD models with different values of black hole mass and Eddington ratio. We adopt the grid shown in Figure \ref{fig:NXSV} with 16 black hole mass and Eddington ratio pairs in the range $\log M_{BH}/M_{\odot}=8-10$ with a step of 0.5\,dex and  $\log\lambda_{Edd}$ in the interval [$-2$, 0] with a logarithmic bin size of 0.5\,dex. All simulations adopt a fixed mean flux for the mock sources, $\log f_X/\rm (erg\,s^{-1}) = -12.8$, which corresponds to an average of $10\pm5$ photon counts per epoch. For each pair of black hole mass and Eddington ratio we generate a total of $N_S = 500$ simulated time series, each with 5 epochs and a cadence of 6 months as explained above. The inferred ensemble NEV of the population is then compared with the intrinsic variance of the input light curves. For the latter we use the Normalised Intrinsic Variance \citep[NIV,][]{Bogensberger2024}. This is measured via Equation \ref{eq:niv} using the simulated light curve segments before introducing Poisson noise, i.e. is independent of measurement errors. The NIV is sensitive to the cadence of the light curve. In our analysis the NIV is estimated for segments with a cadence of 1000\,s and duration of 24 months. Because of the stochastic nature of red noise processes \citep{Vaughan2003} segments generated from the same PSD with fixed cadence have different NIVs. We adopt the median of the distribution as a measure of the intrinsic variance for a given input PSD. 


\subsection{Testing and validation results}\label{ap:results}

\begin{figure}
	\includegraphics[width=1.0\columnwidth]{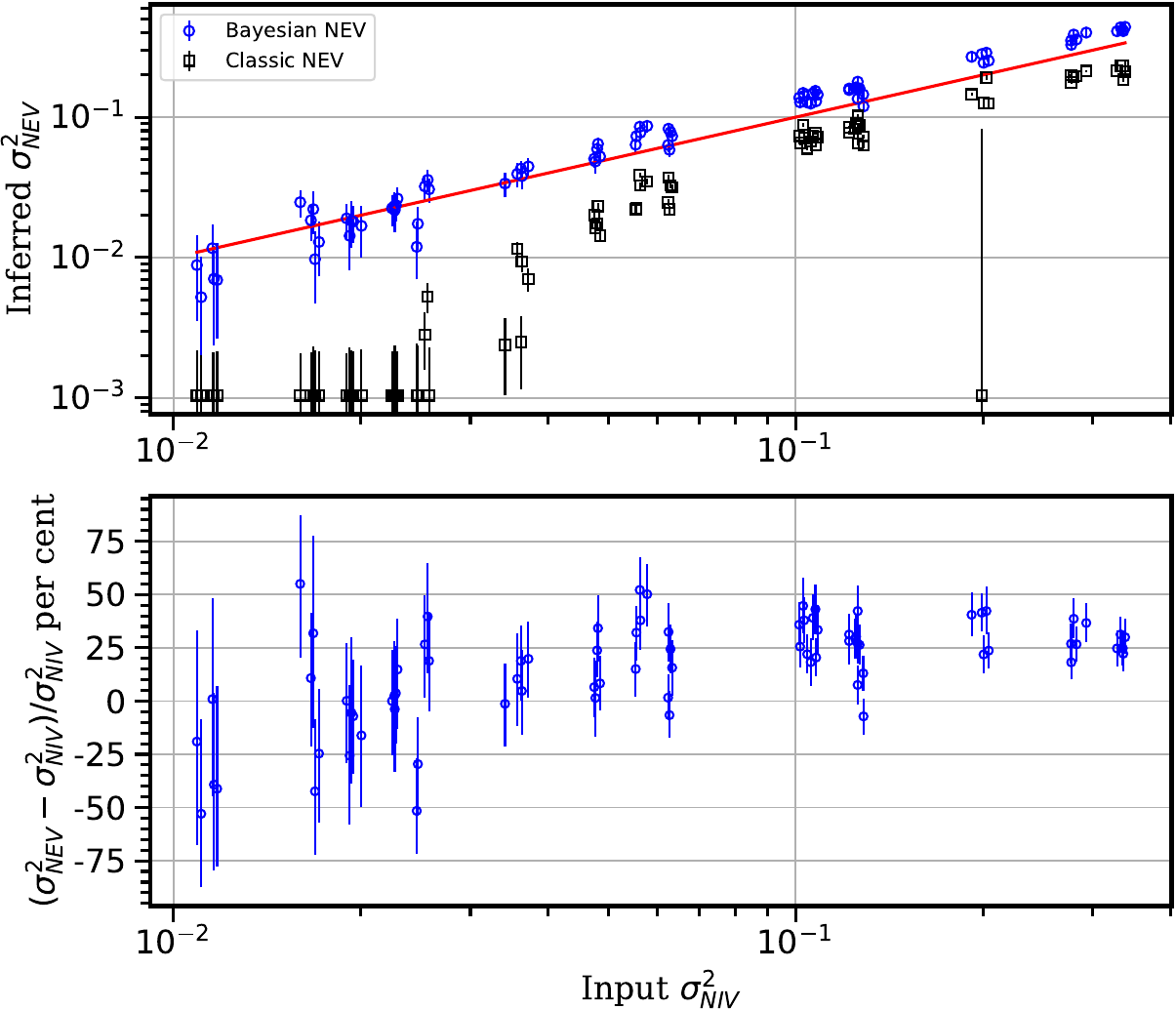}
    \caption{The top panel plots the inferred ensemble NEV vs the input median NIV of the population. The blue circles correspond to the ensemble NEV estimated by the Bayesian approach of section \ref{sec:method}. The error-bars mark the 68\% confidence interval around the median. Each data point corresponds to the ensemble of a total of 500 simulated sources, each of which is assigned an X-ray flux time-series of 5 epochs with a cadence of 6 months. The black squares show the ensemble NEV estimated by applying  Equation \ref{eq:nev} to each of the 500 time-series and then estimating the median and standard error of the population. Negative ensemble NEV values are plotted at $\sigma^2_{NEV}=10^{-3}$. The simulated light curves have been generated for 16 different PSD as explained in the text using 5 realisations per PSD. The bottom panel shows the fractional percentage difference between the inferred NEV and and input median NIV. Only the  {\sc eBExVar}  estimates are shown in this panel.}
    \label{fig:nxsv_validation_NS500_C5}
\end{figure}

\begin{figure}
	\includegraphics[width=1.0\columnwidth]{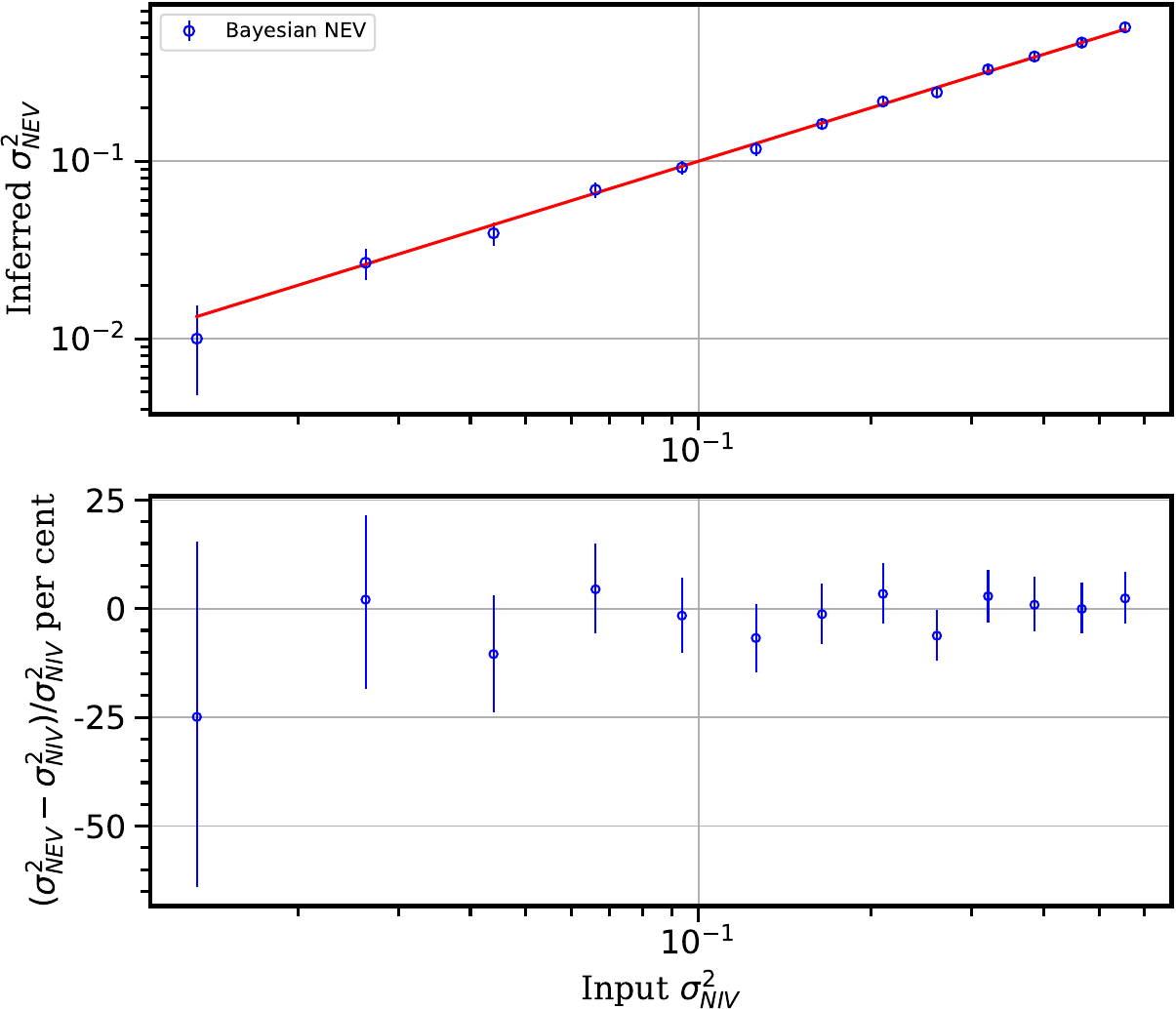}
    \caption{The top panel plots the inferred ensemble NEV vs input NIV in the case of white-noise light curves generated by drawing fluxes from a log-normal distribution with scatter parameter $\sigma_{NIV}$. The blue circles correspond to the ensemble NEV estimated by the Bayesian approach of section \ref{sec:method}. The error-bars mark the 68\% confidence interval around the median. Each data point corresponds to the ensemble of a total of 500 simulated sources, each of which is assigned an X-ray flux time-series of 5 epochs with a cadence of 6 months. The bottom panel shows the fractional percentage difference between the inferred $\sigma^2_{NEV}$ and input $\sigma^2_{NIV}$.}
    \label{fig:nxsv_validation_NS500_C5_wnoise}
\end{figure}

\begin{figure}
	\includegraphics[width=1.0\columnwidth]{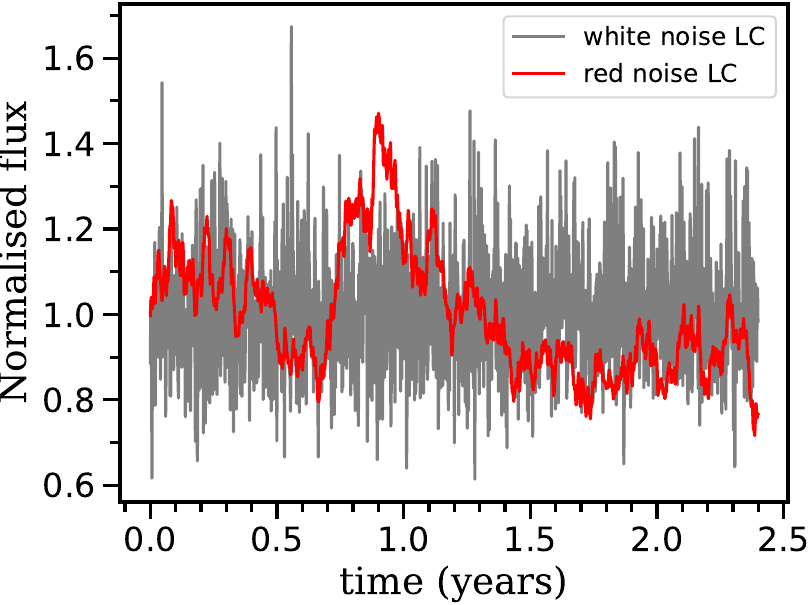}
    \caption{Examples of light curves with the same variance generated via either a white-noise process (grey) or a red-noise PSD (red lines). For the latter we adopt the bending power-law of Equation \ref{eq:psd} with a break frequency given by Equation \ref{eq:model1-nub} and amplitude that follows Equation \ref{eq:model3-amp}. The black hole mass is set to $\log M_{BH}/M_{\odot}=9.75$ and the Eddington ratio is fixed to $\log \lambda_{Edd}=-0.25$.}
    \label{fig:example_lcs}
\end{figure}

\begin{figure}
	\includegraphics[width=1.0\columnwidth]{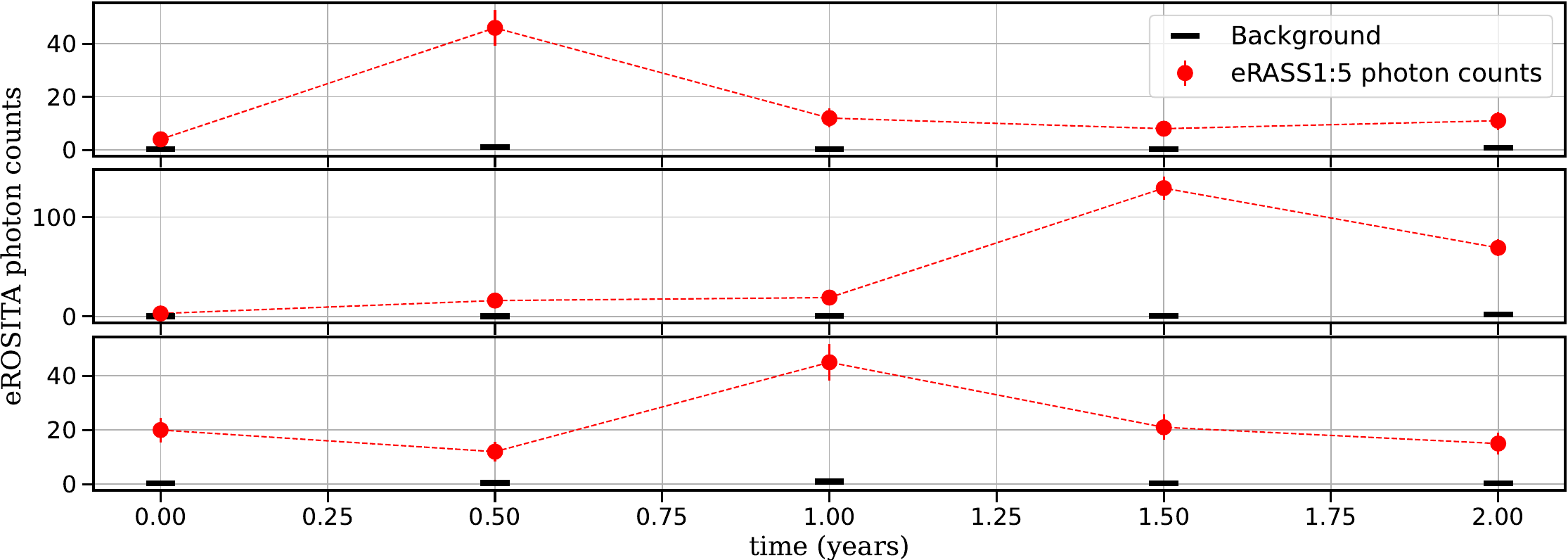}
    \caption{Examples of simulated eROSITA photon count light curves with 5-epoch observations separated by 6\,months. The light curves are chosen to demonstrate erratic behaviour, i.e. they contain one epoch at which the simulated photon counts significantly deviate from the median (difference of more than 20 photon counts). The simulated light curves are generated from the bending power-law of Equation \ref{eq:psd} with a break frequency given by Equation \ref{eq:model1-nub} and amplitude that follows Equation \ref{eq:model3-amp}. The black hole mass is set to $\log M_{BH}/M_{\odot}=8.75$ and the Eddington ratio is fixed to $\log \lambda_{Edd}=-0.25$.}
    \label{fig:lc_example_erratic}
\end{figure}

Figure \ref{fig:nxsv_validation_NS500_C5} presents the results on the NEV inferred by sampling the likelihood of Equation \ref{eq:like-ensemble}, which assumes log-normal priors for both the single epoch count rates of each source [$\mathcal{Y} ( f_{i,k}\;| \;m_k,\; \sigma_k )$ term in the equation above] and the distribution of NIVs of individual segments [$\mathcal{Y}(\sigma^2_{NIV,k} \;| \;\varSigma,\; B )$ term of Equation \ref{eq:like-ensemble}]. The inferred NEV plotted in Figure \ref{fig:nxsv_validation_NS500_C5} is the median of the log-normal distribution $\mathcal{Y} \left(\varSigma,\; B \right)$ of Equation \ref{eq:like-ensemble}, i.e. $\sigma^2_{NEV}=\rm e^{\varSigma}$. This is compared with the median of the NIV of individual segments used to generate the flux time-series. Figure \ref{fig:nxsv_validation_NS500_C5} shows that the inferred NEV is in reasonable agreement with the input NIV. These results should also be compared with the classic approach of estimating the ensemble NEV that uses Equation \ref{eq:nev} for individual mock sources and assumes that the count rate uncertainty of a given epoch is $\sigma_i=\sqrt{N}$, where $N$ is the observed number of photon counts. The median and standard error of the population are used in this case to approximate the inferred NEV. Figure \ref{fig:nxsv_validation_NS500_C5} demonstrates that unlike  {\sc eBExVar},  the classic approach for estimating the NEV fails to recover the median NIV of the population. 

However, Figure \ref{fig:nxsv_validation_NS500_C5} also shows that in this experimental setup {\sc eBExVar} overestimates by about 25\% the NIV. This systematic is more evident for $\sigma^2_{NIV}\ga 0.1$. It is also not reduced if for example we increase the size of the population that is used to estimate the ensemble variability. Instead we track back this issue to the red-noise nature of the flux variations. This is demonstrated in Figure  \ref{fig:nxsv_validation_NS500_C5_wnoise} which repeats the light curve simulations for a white noise process, whereby the count rates of individual epochs are drawn from  a log-normal distribution. Instead of using a PSD to generate red-noise light curves we use a log-normal random number generator to produce a long light curve  with a given input scatter parameter, $\sigma^2_{NIV}$. This is then split into segments and flux time-series as explained in Section \ref{ap:settingup}. For this setup we find no systematic in the derivation of the NEV from the mock observations. 

The difference between red and white-noise processes is further visualised in Figure \ref{fig:lc_example_erratic} that compares light curves with same variance generated by the two processes. As expected, the variability of the red-noise light curve is clearly correlated with distinct valleys and peaks. Sampling such a time sequence with low cadence observations may produce light curves with large amplitude variations between maximum and minimum flux. Examples of such light curves are shown in Figure \ref{fig:lc_example_erratic}. These are selected from the same simulations used to produce Figure \ref{fig:nxsv_validation_NS500_C5} with the criterion that the maximum photon count is very different from the median (more than 20 photons difference). The determination of the intrinsic properties (NIV, mean flux) of such a light curve is challenging because of the erratic photon count behaviour and the low cadence. It is these type of light curves that are primarily responsible for the 25\% systematic in the determination of the NEV in Figure \ref{fig:nxsv_validation_NS500_C5}. Excluding from the ensemble variability analysis the small fraction (typically few per cent) of the most extreme light curves with the larger difference between maximum and median photon counts eliminates the systematic in Figure \ref{fig:nxsv_validation_NS500_C5}.  Additionally, the higher the intrinsic variance of the red-noise process the more common light curves such as those shown in Figure \ref{fig:lc_example_erratic} are. This explains why the systematic in the NEV determination becomes more evident for $\sigma^2_{NIV}\ga0.1$ in Figure \ref{fig:nxsv_validation_NS500_C5}. 

\begin{figure}
	\includegraphics[width=1.0\columnwidth]{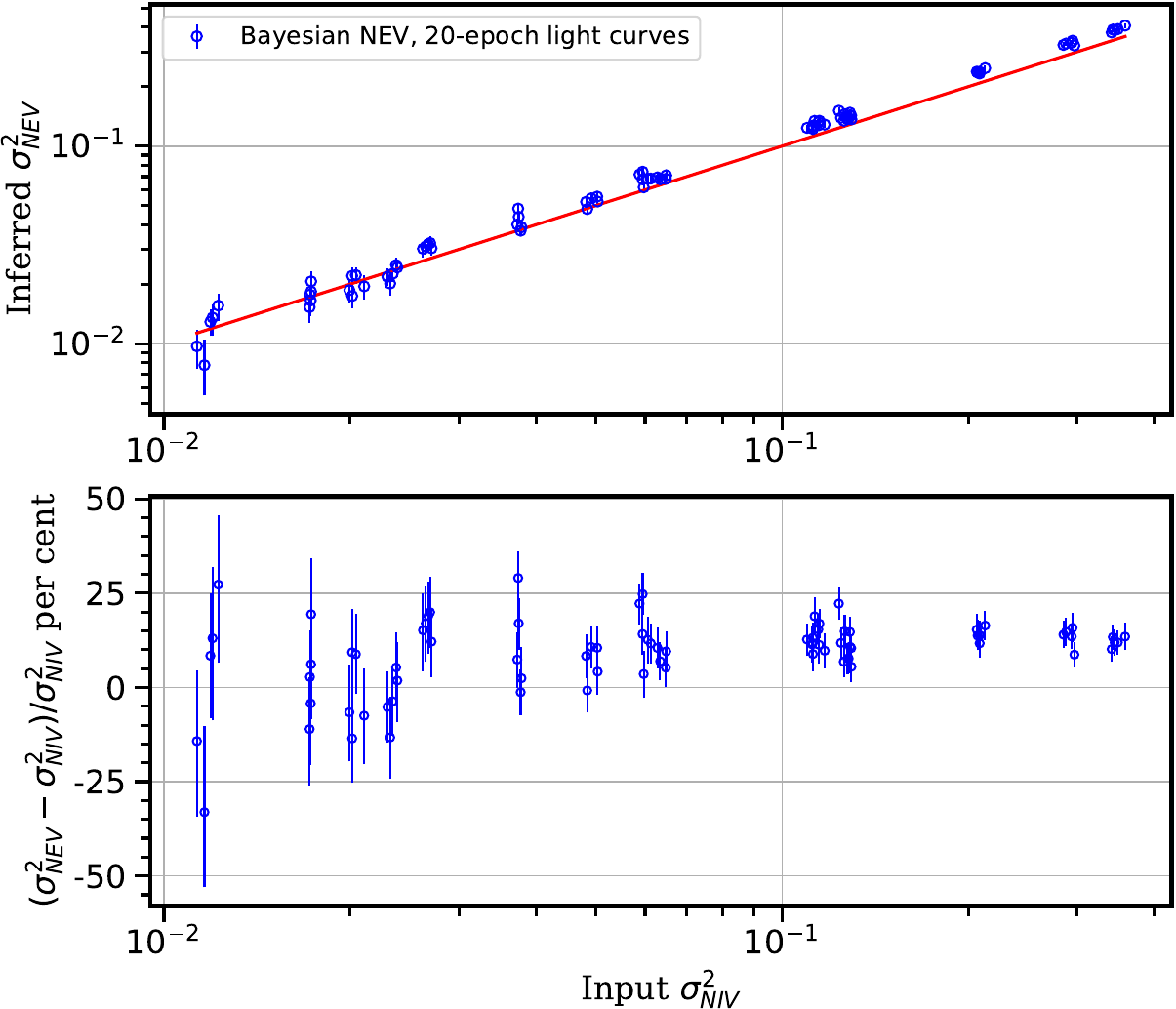}
    \caption{Same as Figure \ref{fig:nxsv_validation_NS500_C5} but for simulated light curves with 20 epoch observations over a period of 2 years instead of 5). The simulated light curves have been generated for 16 different PSD as explained in the text using 5 realisations per PSD. The bottom panel shows the fractional percentage difference between the inferred NEV and and input median NIV. The data points and panels are the same as in Figure \ref{fig:nxsv_validation_NS500_C5}. In both panels only the  {\sc eBExVar} estimates are shown.}
    \label{fig:nxsv_validation_NS500_C20}
\end{figure}

\begin{figure}
	\includegraphics[width=1.0\columnwidth]{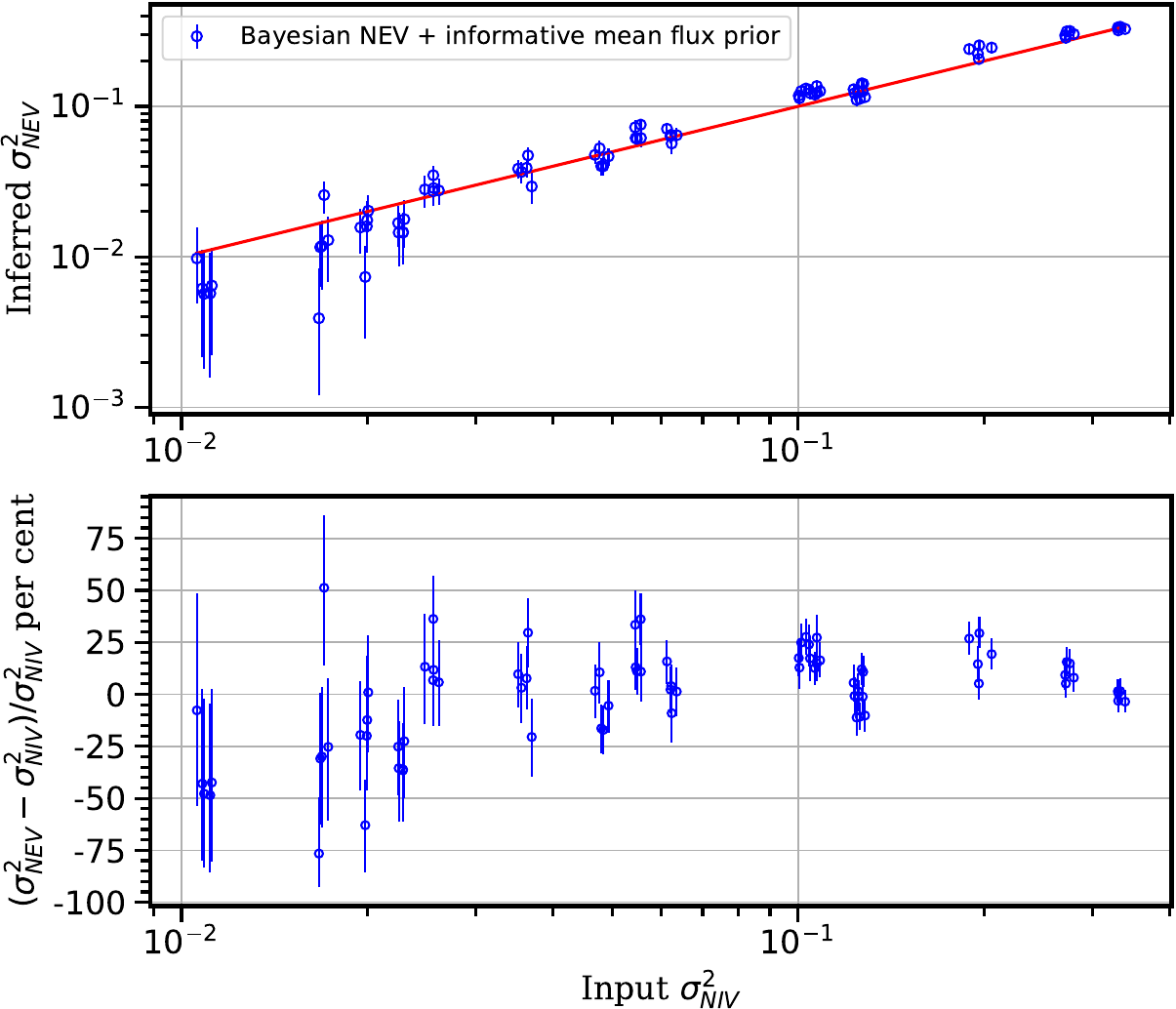}
    \caption{Same as Figure \ref{fig:nxsv_validation_NS500_C5} but including a strong informative prior on the mean flux of the light curve of each source when applying the Bayesian approach of section \ref{sec:method}. The adopted prior is a normal distribution with small scatter (0.2\,dex) and a mean logarihmic flux determined as explained in Appenix \ref{ap:meanflux}. The datapoints and panels are the same as in Figure \ref{fig:nxsv_validation_NS500_C5}. In both panels only the  {\sc eBExVar} estimates are shown.}
    \label{fig:nxsv_validation_NS500_C5nprior}
\end{figure}

We identify two approaches to address the systematic uncertainties in the estimation of the NEV. The first is to improve the observational data by increasing the cadence of the light curves. The second is to place a strong informative prior on the mean flux of the light curve. We have experimented with both approaches using the same diversity of PSDs used in Figure \ref{fig:nxsv_validation_NS500_C5}. For example, Figure \ref{fig:nxsv_validation_NS500_C20} demonstrates that generating light curves with 20 epochs instead of 5 reduces the systematic of the NEV to 15\%. Similarly, Figure \ref{fig:nxsv_validation_NS500_C5nprior} shows the comparison between NEV and NIV in the case of 5-epoch light curves by applying a normal distribution prior to the mean logarithmic count rate of each source with scatter parameter 0.2\,dex and a mean value estimated from the coadded light curve photon counts as explained in Appendix \ref{ap:meanflux}. 
 
\subsection{Estimating the mean flux of a light curve}\label{ap:meanflux}

Suppose an X-ray light curve with $N$ epochs and $N_i$, $B_i$, $t_i$ are respectively the total number of photons within an aperture, the expected background level within the same aperture and the exposure time at the epoch $i$. We estimate the mean count rate of the source by integrating along the time dimension. The probability of a count rate $CR$ for the time-collapsed counts is given by the Poisson form

\begin{equation}\label{eq:flux-poisson}
    P(CR) = \frac{e^{-\lambda} \cdot\, \lambda^{N} } {N!},
\end{equation}

\noindent where $N=\sum_{i=1}^{N} N_i$ is the sum of the source's photons in each epoch.  The Poisson expectation value in the equation above can be written as

\begin{equation}\label{eq_sum_flux}
\lambda = CR \cdot \sum_{i=1}^{N} \bigl( EEF_i \cdot t_i\bigr)+ \sum_i B_i,
\end{equation}

\noindent where $EEF_i$  is the  encircled energy fraction of the photometry aperture at epoch $i$. Equation \ref{eq:flux-poisson} can be solved numerically to determine the X-ray count rate probability distribution function. The mode of this distribution is used to represent the estimate of the mean count-rate of the light curve.  

\subsection{Integration limits for the estimation of \texorpdfstring{$NIV_\infty$}{TEXT}}\label{ap:niv}

The limits for the integration of the PSD to calculate $NIV_\infty$ are directly related to the sampling pattern of the eROSITA light curves, for which the duration of individual time bins is much smaller than the interval between successive observations. The scan rate of SRG/eROSITA is 90 degrees per hour. It therefore completes a great circle every approximately 4\,hours. This time interval is referred to as one eroday. A source can be observed for up to about 41\,s per eroday as it moves across the eROSITA field of view. Additionally, the direction of the angular momentum vector of SRG/eROSITA shifts by an average of 10\,arcmin per eroday along the ecliptic plane in order to complete a full sky survey in a period of 6 months. This shift means that a given position on the sky is revisited approximately every 4\,hours for a total of typically 6 consecutive erodays. The accumulated exposure time per sky position during these consecutive scans corresponds to a single eROSITA All Sky Survey. Therefore the total exposure time of a source per eRASS  is accumulated over a period of about 24\,hours and successive eRASS observations are separated by about 6 months. This sampling pattern is very different from a setup where successive time bins of a light curve are adjacent with no gaps between them. For the eROSITA light curves power from frequencies larger than the sampling frequency affects the average count rate of each time bin \citep[e.g.][]{Bogensberger2024}. 
 
Therefore when comparing our results with the $NIV_\infty$ estimated by integrating model PSDs (e.g. Figure \ref{fig:NXSV}), the relevant maximum integration frequency corresponds to the timescale $24 / (1+z)$\,hours, where $z$ is the redshift of the source. Although the eRASS1-5 light curves have a cadence of 6\,months, we are not averaging the source flux on that timescale. Instead the eRASS1-5 observations retain information on the variability of a source on timescales $<6$\,months. This point is already implied by the comparisons shown in Figures \ref{fig:nxsv_validation_NS500_C5} and  \ref{fig:nxsv_validation_NS500_C5nprior}. Nevertheless, we also use dedicated simulations to quantify the argument above.

We use the same mock light curves for a given input model PSD described in Appendix \ref{ap:settingup} with a cadence of 1,000\,s. We then resample them to a cadence 24\,hours by averaging the mock light curve fluxes within that time interval. As a result any variability information on times scales shorter than 24\,h is lost. For each resampled light curve we then select 400 segments with a duration of 2\,years and a cadence of 6\,months. This translates to a total of 5 photometric points per segment with an effective exposure time of 24\,h per point. For this exercise no noise is added to the light curve. 

\begin{figure}
	\includegraphics[width=\columnwidth]{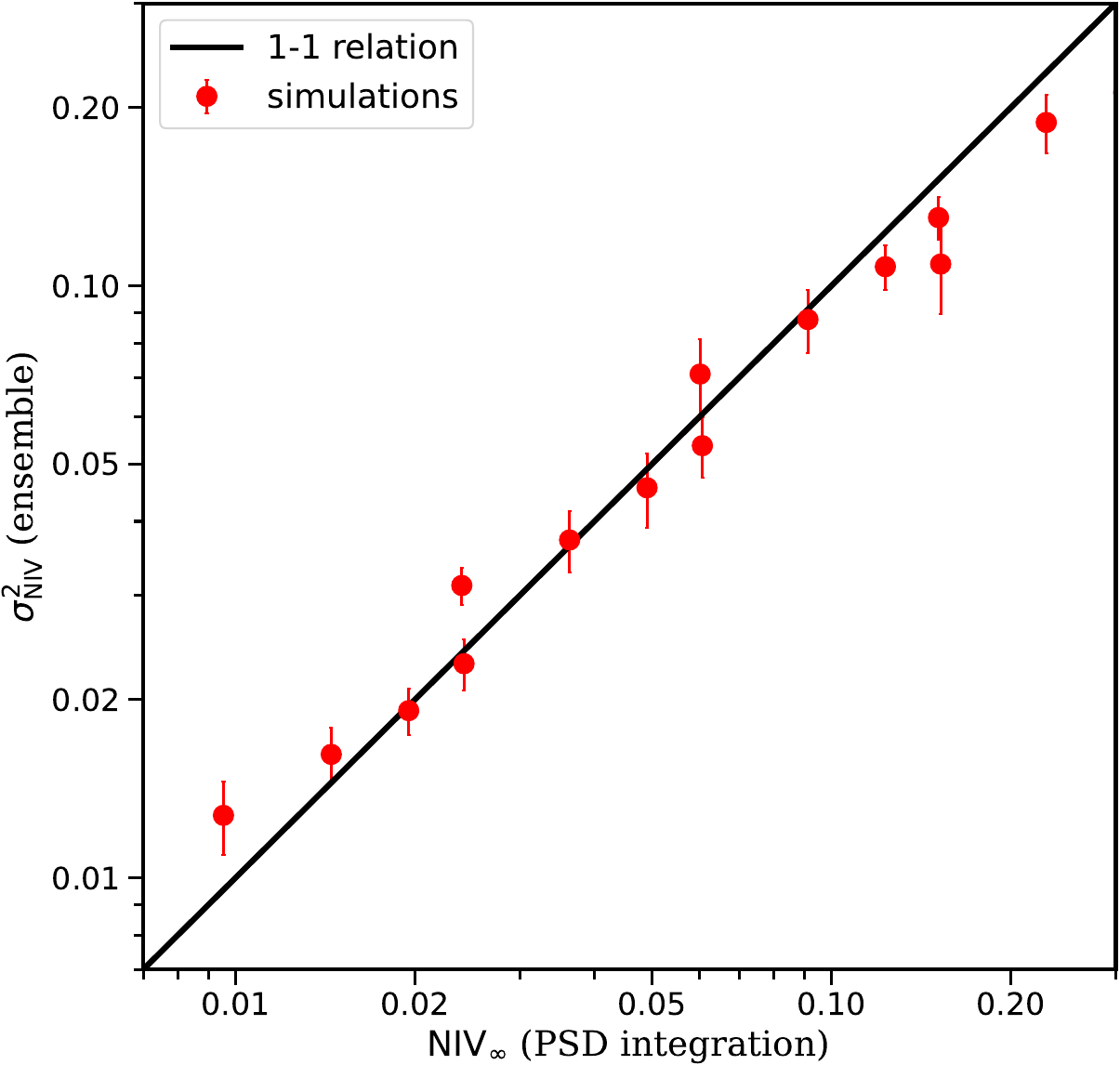}
    \caption{Comparison between the ensemble NIV (y--axis) of the simulated light curve segments described in Section \ref{ap:niv} with the $NIV_\infty$ (x--axis) estimated from the PSD integration as discussed in the same section. Each data point on this plot corresponds to a PSD with fixed black hole mass and Eddington ratio (Equations \ref{eq:psd}-\ref{eq:model3-amp}). For each PSD we generate a total of 10 realisations of the 400 light curve segments described in Appendix \ref{ap:niv}. The data points show the average and standard deviation of the 10 independent realisations. The black line shows the one--to--one relation.}
    \label{fig:NIV_NIVINF}
\end{figure}

We then estimate the normalised intrinsic variance of each segment using Equation \ref{eq:niv}. These variances are then used to estimate the ensemble, i.e. the average of the 400 segments. This can be compared with the $NIV_\infty$ estimated by directly integrating the PSD. For this latter calculation we use the frequencies that correspond to the timescales $\rm \Delta t$=24\,h (i.e. the effective exposure time of each photometric point of the segment, below which any variability information is lost) and 2\,yr (duration of segment). The results of this comparison are shown in Figure \ref{fig:NIV_NIVINF}. What this plot demonstrates is that for a light curve sampling pattern similar to that of eROSITA (one photometric observation accumulated over a few thousand seconds every 6 months for a period of 2 years) it is the bin duration and not the light curve cadence that is relevant for the calculation of $NIV_\infty$. 

The results above are identical to those of \cite{Allevato2013} who studied in detail the statistical properties of the excess variance as measured by sparse but uniformly sampled light curves. They also found that the main property of a light curve segment that guarantees the unbiased measurement of its variance is the uniform sampling, and not so much the number of the points in it.

\section{NEV Analytic fit}\label{ap:polynomial}

We use a 2-dimensional polynomial to fit the measured NEV of QSOs presented in Figure \ref{fig:NXSV} as function of black hole mass and Eddington ratio. The analytic model has the form

\begin{equation}\label{eq:polynomial}
    \sigma_{NEV}^2 = \sum_{i,j\le2}\,A_{i,j}\cdot \left[ \log \frac{M_{BH}}{10^8\,M_\odot} \right]^i \cdot \left[ \log \frac{\lambda_{Edd}}{10^{-2}} \right]^j .
\end{equation}

\begin{figure}
	\includegraphics[width=\columnwidth]{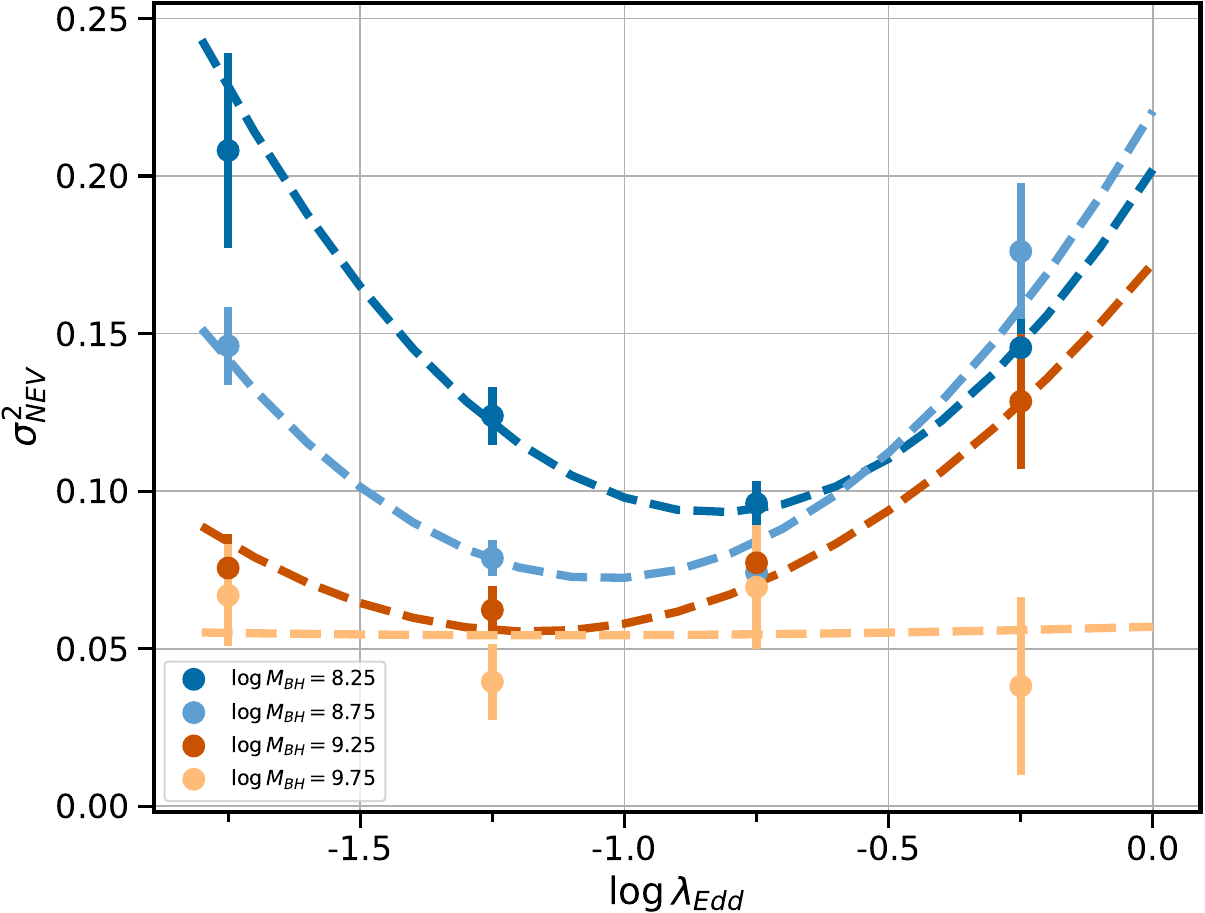}
    \caption{NEV as a function of Eddington ratio. The data points correspond to the measurements presented in Figure \ref{fig:NXSV} for different black hole mass bins as indicated by the different colours. The dashed curves show the analytic fit to the data points using Equation \ref{eq:polynomial}. The colour coding of the curves is the same as that of the data points.}
    \label{fig:polynomial}
\end{figure}

\noindent The coefficients are estimated via least-square fit to the observations and are given in Table   \ref{tab:polynomial}. Figure  \ref{fig:polynomial} compares the NEV measurements with the analytic fit. We caution that the polynomial fit should not be extrapolated outside the grid of black hole mass and Eddington ratio of Figure \ref{fig:NXSV}.

\begin{table}
    \caption{Coefficients of the analytic fit in Equation \ref{eq:polynomial}}
    \label{tab:polynomial}
    \centering
    \begin{tabular}{l c}
    \hline
        $A_{0,0}$ & $+0.38\pm0.07$\\
        $A_{1,0}$ & $-0.28\pm0.13$ \\
        $A_{2,0}$ & $0.06\pm0.06$ \\
        $A_{0,1}$ & $-0.42\pm0.19$\\
        $A_{1,1}$ & $+0.19\pm0.27$ \\
        $A_{2,1}$ & $+0.03\pm0.13$ \\
        $A_{0,2}$ & $+0.16\pm0.06$\\
        $A_{1,2}$ & $+0.02\pm0.13$ \\
        $A_{2,2}$ & $-0.06\pm0.06$ \\
        \hline
    \end{tabular}
\end{table}

\section{The sensitivity of the inferred NEV on the definition of the DRQ16 sample}\label{ap:validation2}

This section explores the sensitivity of the ensemble excess variance results presented in Figure \ref{fig:NXSV} to the selection details of the QSO sample. We investigate if systematics in e.g. black hole mass estimates or observational effects such as the adopted X-ray spectral band, change the trends shown in Figure \ref{fig:NXSV}.  

\begin{figure}
	\includegraphics[width=\columnwidth]{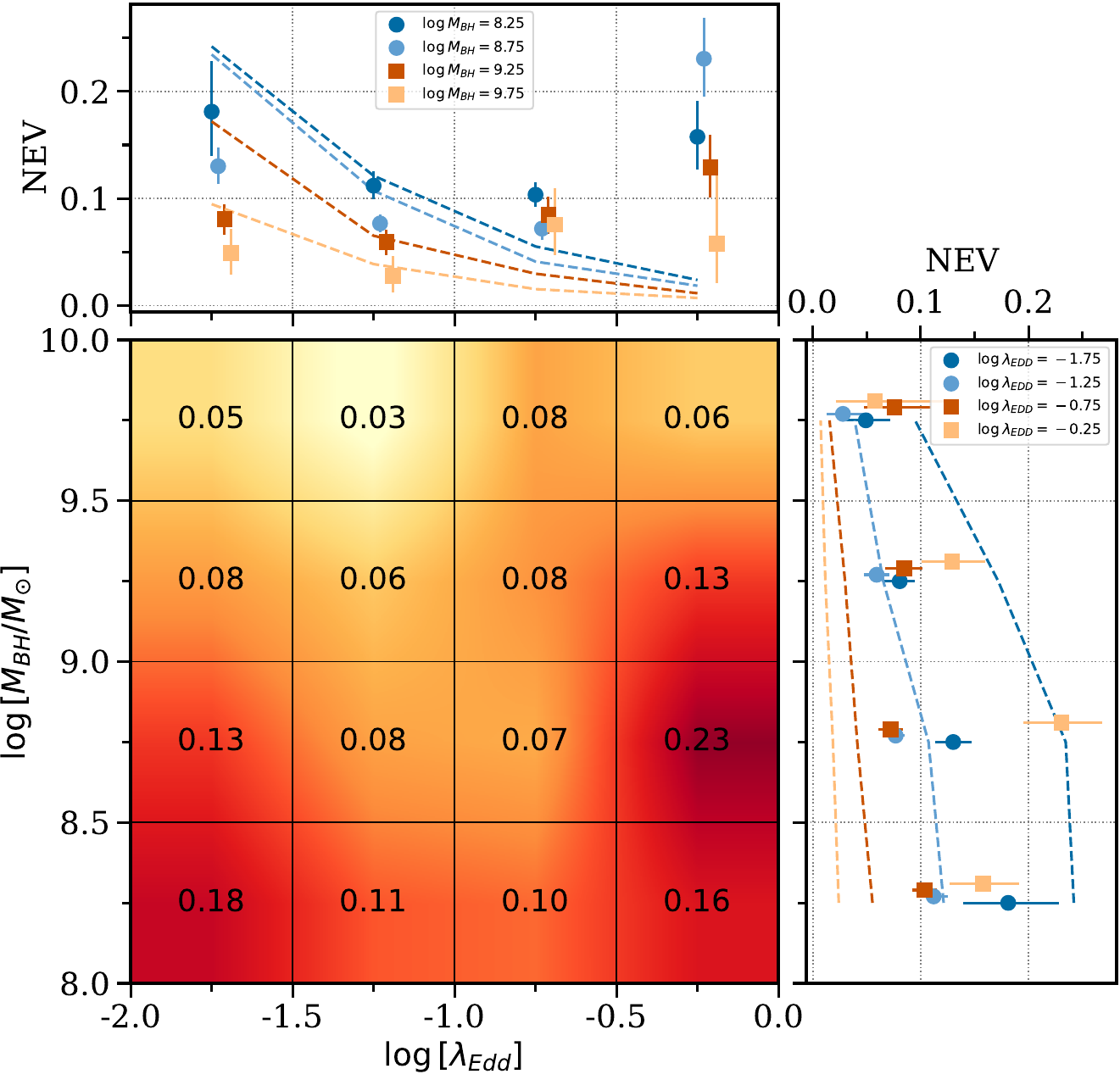}
    \caption{Same as Figure \ref{fig:NXSV} but for the ensemble normalised excess variance of DRQ16 QSOs in the 0.6-2.3\,keV spectral band. The symbols, lines and labels are the same as in  Figure \ref{fig:NXSV}.}
    \label{fig:NXSV_022}
\end{figure}

\subsection{X-ray spectral band}

eROSITA is primarily a soft X-ray mission, with the sensitivity dropping substantially beyond about 2.5\,keV. Therefore there is limited flexibility in exploring variability statistics as a function of energy. Nevertheless, we repeat the analysis of Section \ref{sec:results} for the 0.6-2.3\,keV energy range (eROSITA spectral band 2). X-ray photometry in the 0.6-2.3\,keV  band is derived for the eRASS1 to 5 observations at the positions of DRQ16 QSOs. The selection criteria described in Section \ref{sec:observations} are then applied to produce a  sample of 4389 unique light curves with at least 10 photons in at least one eRASS epoch. The ensemble NEV of this sample is estimated as a function of black hole mass and Eddington ratio and the results are shown in Figure \ref{fig:NXSV_022}. The trends with $M_{BH}$ and $\lambda_{Edd}$ are qualitative similar to those derived for the spectral band 0.2-2.3\,keV and graphically shown in Figure \ref{fig:NXSV}.

\subsection{Black hole mass quality flags}

The baseline sample adopted in this work does not exclude QSOs that may have problematic black hole mass measurements in the catalog of \cite{Wu_Shen2022}. These may introduce systematics in the analysis and bias the ensemble variability results of Figure \ref{fig:NXSV}. We test this by applying to the baseline sample described in Section \ref{sec:observations} the quality cuts proposed by \citet[][see their Section 4]{Wu_Shen2022} to produce clean sample of black hole measurements. The sample size numbers a total of 8602 unique light curves.  The ensemble NEV of this sample is estimated in the same black hole mass and Eddington ratio bins shown in Figure \ref{fig:NXSV}. For the new sample we estimate very similar, within the uncertainties, ensemble NEV as those in Figure \ref{fig:NXSV}. Our results are therefore robust to the quality of the black hole measurements. For the sake of brevity we do not graphically show the ensemble NEV for the new sample. 

\subsection{Sloan QSO selection flags}

The DRQ16 catalogue includes QSOs targeted by the various SDSS spectroscopic programmes (SDSS-I to IV) following diverse selection criteria \citep[e.g.][]{Schneider2010, Ross2012, Myers2015}. In addition to the core QSO samples selected on the basis of their UV/optical and/or mid-infrared colours, there are numerous complementary (filler) SDSS spectroscopic programmes that target AGN/QSOs selected on the basis of their specific properties, such as radio emission, association with X-ray sources, red mid-infrared colours, optical variability, high redshift candidates etc. The baseline sample used in this paper includes all these diverse target groups although it is dominated by the colour-selected QSOs \citep[i.e. \texttt{CORE} selection flags of SDSS-III/BOSS and SDSS-IV/eBOSS,][]{Myers2015, Dawson2016}. 

We explore the sensitivity of our results to the definition of the QSO sample by applying the variability analysis only to colour-selected QSOs. These are the ones that define the \texttt{CORE} samples of SDSS-III/BOSS and SDSS-IV/eBOSS \citep{Myers2015}, including any sources that were already targeted by the earlier SDSS-I and II programmes \citep{Schneider2010}.  

Applying the \texttt{CORE}  selection of the baseline sample described in Section \ref{sec:observations} results to a 7439 unique light curves with multi epoch eRASS photometry in the 0.2-2.3\,keV band.  The ensemble NEV of this new sample is estimated in the same black hole mass and Eddington ratio bins shown in Figure \ref{fig:NXSV}. For the new sample we estimate very similar, within the uncertainties, ensemble NEV as those in Figure \ref{fig:NXSV}. For the sake of brevity we do not graphically show the ensemble NEV for the new sample. 

\subsection{Redshift selected subsamples}

There are strong covariances between black hole mass, Eddington ratio and redshift as a result of the SDSS optical flux limits, QSO selection biases and volume effects. Smaller black hole masses (and hence higher Eddington ratios) tend to be found at lower redshift, whereas more massive black holes (and lower Eddington ratios) are more common at higher redshift. Moreover, redshift variations affect our analysis in two ways. First the rest-frame frequencies probed by the eROSITA light curves differ. Secondly, the rest-frame X-ray band moves to harder energies with increasing redshift. Both effects above may introduce systematics in the determination of the ensemble variability of QSOs, which we investigate in this section. 

The baseline SDSS QSO sample is split at $z=0.8$ into two nearly equal size subsets (5812 for $z<0.8$; 3866 for $z>0.8$). The ensemble NEV is estimated in the same black hole mass and Eddington ratio bins shown in Figure \ref{fig:NXSV}. Overall we observe the same trends as those shown in Figure \ref{fig:NXSV}, although the uncertainties are larger because of the smaller size of the redshift selected subsamples. We conclude that redshift effects do not affect our results and conclusions. 


\bsp	
\label{lastpage}
\end{document}